\documentclass{amsart}

\usepackage{amsmath}
\usepackage{amsfonts}
\usepackage{amssymb}

\usepackage{geometry}
\usepackage{graphics}
\usepackage[dvips]{graphicx}
\usepackage{pgf}

\usepackage{mathrsfs}
\usepackage{tabularx}

\usepackage{amssymb}
\usepackage{tikz}
\usetikzlibrary{arrows,automata,positioning}

\tikzstyle{vertex}=[circle, draw, inner sep=0pt, minimum size=4pt]

\usepackage{xcolor,soul}

\usetikzlibrary{arrows,shapes,decorations,automata,backgrounds,petri,positioning}

\tikzset{main node/.style={circle,fill=blue!20,draw,minimum size=1cm,inner sep=0pt},
            }

\usepackage{amsthm}
\usepackage{bbm}
\usepackage{MnSymbol}	
\usepackage{color}
\usepackage{mathrsfs}

\def\qed{\begin{flushright} $\square$ \end{flushright}}
\def\qee{\begin{flushright} $\Diamond$ \end{flushright}}

\catcode`\@=11
\long\def\@makefntext#1{
\protect\noindent \hbox to 3.2pt {\hskip-.9pt  
$^{{\eightrm\@thefnmark}}$\hfil}#1\hfill}		

\def\@makefnmark{\hbox to 0pt{$^{\@thefnmark}$\hss}}	
	
\def\ps@myheadings{\let\@mkboth\@gobbletwo
\def\@oddhead{\hbox{}
\rightmark\hfil\eightrm\thepage}   
\def\@oddfoot{}\def\@evenhead{\eightrm\thepage\hfil
\leftmark\hbox{}}\def\@evenfoot{}
\def\sectionmark##1{}\def\subsectionmark##1{}}

\oddsidemargin=\evensidemargin
\newcounter{sectionc}\newcounter{subsectionc}\newcounter{subsubsectionc}
\renewcommand{\section}[1] {\vspace{12pt}\addtocounter{sectionc}{1} 
\setcounter{subsectionc}{0}\setcounter{subsubsectionc}{0}\noindent 
	{\tenbf\thesectionc. #1}\par\vspace{5pt}}
\renewcommand{\subsection}[1] {\vspace{12pt}\addtocounter{subsectionc}{1} 
\setcounter{subsubsectionc}{0}\noindent 
{\bf\thesectionc.\thesubsectionc. {\kern1pt \bfit #1}}\par\vspace{5pt}}
\renewcommand{\subsubsection}[1] {\vspace{12pt}\addtocounter{subsubsectionc}{1}
	\noindent{\tenrm\thesectionc.\thesubsectionc.\thesubsubsectionc.
	{\kern1pt \tenit #1}}\par\vspace{5pt}}
\newcommand{\nonumsection}[1] {\vspace{12pt}\noindent{\tenbf #1}
	\par\vspace{5pt}}

\newcounter{appendixc}
\newcounter{subappendixc}[appendixc]
\newcounter{subsubappendixc}[subappendixc]
\renewcommand{\thesubappendixc}{\Alph{appendixc}.\arabic{subappendixc}}
\renewcommand{\thesubsubappendixc}
	{\Alph{appendixc}.\arabic{subappendixc}.\arabic{subsubappendixc}}

\renewcommand{\appendix}[1] {\vspace{12pt}
        \refstepcounter{appendixc}
        \setcounter{figure}{0}
        \setcounter{table}{0}
        \setcounter{lemma}{0}
        \setcounter{theorem}{0}
        \setcounter{corollary}{0}
        \setcounter{definition}{0}
        \setcounter{equation}{0}
        \renewcommand{\thefigure}{\Alph{appendixc}.\arabic{figure}}
        \renewcommand{\thetable}{\Alph{appendixc}.\arabic{table}}
        \renewcommand{\theappendixc}{\Alph{appendixc}}
        \renewcommand{\thelemma}{\Alph{appendixc}.\arabic{lemma}}
        \renewcommand{\thetheorem}{\Alph{appendixc}.\arabic{theorem}}
        \renewcommand{\thedefinition}{\Alph{appendixc}.\arabic{definition}}
        \renewcommand{\thecorollary}{\Alph{appendixc}.\arabic{corollary}}
        \renewcommand{\theequation}{\Alph{appendixc}.\arabic{equation}}
        \noindent{\tenbf Appendix \theappendixc #1}\par\vspace{5pt}}
\newcommand{\subappendix}[1] {\vspace{12pt}
        \refstepcounter{subappendixc}
        \noindent{\bf Appendix \thesubappendixc. {\kern1pt \bfit #1}}
	\par\vspace{5pt}}
\newcommand{\subsubappendix}[1] {\vspace{12pt}
        \refstepcounter{subsubappendixc}
        \noindent{\rm Appendix \thesubsubappendixc. {\kern1pt \tenit #1}}
	\par\vspace{5pt}}

\newcommand{\textlineskip}{\baselineskip=13pt}
\newcommand{\smalllineskip}{\baselineskip=10pt}

\def\abstracts#1#2#3{{
	\centering{\begin{minipage}{4.5in}\footnotesize\baselineskip=10pt
	\parindent=0pt #1\par 
	\parindent=15pt #2\par
	\parindent=15pt #3
	\end{minipage}}\par}} 

\def\keywords#1{{
	\centering{\begin{minipage}{4.5in}\footnotesize\baselineskip=10pt
	{\footnotesize\it Keywords}\/: #1
	 \end{minipage}}\par}}

\renewenvironment{thebibliography}[1]
        {\frenchspacing
	 \ninerm\baselineskip=11pt
         \begin{list}{\arabic{enumi}.}
        {\usecounter{enumi}\setlength{\parsep}{0pt}     
	 \setlength{\leftmargin 12.7pt}{\rightmargin 0pt}
         \setlength{\itemsep}{0pt} \settowidth
	{\labelwidth}{#1.}\sloppy}}{\end{list}}

\newcounter{itemlistc}
\newcounter{romanlistc}
\newcounter{alphlistc}
\newcounter{arabiclistc}

\newcommand{\fcaption}[1]{
        \refstepcounter{figure}
        \setbox\@tempboxa = \hbox{\footnotesize Fig.~\thefigure. #1}
        \ifdim \wd\@tempboxa > 5in
           {\begin{center}
        \parbox{5in}{\footnotesize\smalllineskip Fig.~\thefigure. #1}
            \end{center}}
        \else
             {\begin{center}
             {\footnotesize Fig.~\thefigure. #1}
              \end{center}}
        \fi}

\newcommand{\tcaption}[1]{
        \refstepcounter{table}
        \setbox\@tempboxa = \hbox{\footnotesize Table~\thetable. #1}
        \ifdim \wd\@tempboxa > 5in
           {\begin{center}
        \parbox{5in}{\footnotesize\smalllineskip Table~\thetable. #1}
            \end{center}}
        \else
             {\begin{center}
             {\footnotesize Table~\thetable. #1}
              \end{center}}
        \fi}

\def\pmb#1{\setbox0=\hbox{#1}
	\kern-.025em\copy0\kern-\wd0
	\kern.05em\copy0\kern-\wd0
	\kern-.025em\raise.0433em\box0}

\def\fnt#1#2{\footnotetext{\kern-.3em
	{$^{\mbox{\scriptsize #1}}$}{#2}}}

\def\fpage#1{\begingroup
\voffset=.3in
\thispagestyle{empty}\begin{table}[b]\centerline{\footnotesize #1}
	\end{table}\endgroup}

\font\tenrm=cmr10
\font\tenit=cmti10 
\font\tenbf=cmbx10
\font\bfit=cmbxti10 at 10pt
\font\ninerm=cmr9

\font\eightrm=cmr8

\newtheorem{theorem}{\indent Theorem}

\newtheorem{lemma}[theorem]{Lemma}

\newtheorem{example}{Example}

\newtheorem{pro}[theorem]{Proposition}

\newtheorem{cor}[theorem]{Corollary}
\newtheorem{remark}{Remark}

\def\FigName{figure}%
\newbox\captionbox
\long\def\@makecaption#1#2{%
  \ifx\FigName\@captype
    \vskip\abovecaptionskip
    \setbox\tempbox\hbox{{\figurecaptionfont #1\hskip1em #2}}
	\ifdim\wd\tempbox< 28pc
	\centerline{\box\tempbox}
	\else
	{\figurecaptionfont #1\hskip1em #2\par}
\fi\else
  	\setbox\tempbox\hbox{{\tablecaptionfont #1\hskip1em #2}}
 	\ifdim\wd\tempbox< 28pc 
	\centerline{\box\tempbox}
	\else
	{\tablecaptionfont #1\hskip1em #2\par}%
	\fi   
 \vskip\belowcaptionskip
 \fi}
\InputIfFileExists{psfig.sty}
{\typeout{^^Jpsfig.sty inputed...ok}}{\typeout{^^JWarning: psfig.sty could be be found.^^J}}
\InputIfFileExists{epsfsafe.tex}
{\typeout{^^Jepsfsafe.tex inputed...ok}}
			{\typeout{^^JWarning: epsfsafe.tex could not be found.^^J}}
\InputIfFileExists{epsfig.sty}
{\typeout{^^Jepsfig.sty inputed...ok}}{\typeout{^^JWarning: epsfig.sty could not be found.^^J}}
\InputIfFileExists{epsf.sty}
{\typeout{^^Jepsf.sty inputed...ok}}{\typeout{^^JWarning: epsf.sty could not be found.^^J}}%
\def\fps@figure{tbp}
\def\ftype@figure{1}
\def\ext@figure{lof}
\def\fnum@figure{Fig.\ \thefigure}
\def\qed{\hbox{${\vcenter{\vbox{	          
   \hrule height 0.4pt\hbox{\vrule width 0.4pt height 6pt
   \kern5pt\vrule width 0.4pt}\hrule height 0.4pt}}}$}}

\begin{document}

\def\beq{\begin{equation}}
\def\eeq{\end{equation}}
\def\cqfd{\begin{flushright} $\square$ \end{flushright}}
\def\cqfde{\begin{flushright} $\Diamond$ \end{flushright}}
\def\bex{\begin{example}}
\def\eex{\end{example}}



\normalsize\textlineskip
\thispagestyle{empty}
\setcounter{page}{1}



\fpage{1}
\centerline{\bf
HOMOGENEOUS OPEN QUANTUM WALKS ON THE LINE: }
\vspace*{0.035truein}
\centerline{\bf  CRITERIA FOR SITE RECURRENCE AND ABSORPTION
}
\vspace*{0.37truein}
\centerline{\footnotesize 
T. S. JACQ AND C. F. LARDIZABAL\footnote{Emails: tjacq@ufrgs.br, cfelipe@mat.ufrgs.br}}
\vspace*{0.015truein}
\centerline{\footnotesize\it Instituto de Matem\'atica e Estat\'istica - Universidade Federal do Rio Grande do Sul - UFRGS}
\baselineskip=10pt
\centerline{\footnotesize\it Av. Bento Gon\c calves 9500 CEP 91509-900, Porto Alegre, RS, Brazil
}
\vspace*{10pt}


\vspace*{0.21truein}
\abstracts{
In this work, we study open quantum random walks,  as described by S. Attal et al.\cite{attal}.
These objects are given in terms of completely positive maps acting on trace-class operators, leading to one of the simplest open quantum versions of the recurrence problem for classical, discrete-time random walks. This work focuses on obtaining criteria for site recurrence of nearest-neighbor, homogeneous walks on the integer line, with the description presented here making use of recent results of the theory of open walks, most particularly regarding reducibility properties  of the operators involved. This allows us to obtain a complete criterion for site recurrence in the case for which the internal degree of freedom of each site (coin space) is of dimension 2. We also present the analogous result for irreducible walks with an internal degree of arbitrary finite dimension and the absorption problem for walks on the semi-infinite line.
}{}{}

\vspace*{10pt}
\vspace*{3pt}

\vspace*{1pt}\textlineskip	
\section{Introduction}	        
\vspace*{-0.5pt}
\noindent


One of the most basic results in the classical theory of Markov chains regards the discrete-time random walk on the integer line for which a particle moves left with probability $p$, and moves right with probability $1-p$ \cite{durrett,grimmett}. In this setting, we have the recurrence problem: the probability of eventually returning to the origin equals 1 if, and only if, the coin is fair, that is, if $p=1/2$. The focus of this work is on the study of the recurrence problem in an open quantum setting: if a quantum particle is located on a certain position $i$ of the line with an internal degree of freedom described by a finite dimensional density matrix $\rho_i$, then we define one step of a nearest-neighbor walk as
\beq\label{bas1} \rho_i\otimes |i\rangle\langle i| \;\mapsto\; L\rho_i L^*\otimes|i-1\rangle\langle i-1|+R\rho_i R^*\otimes|i+1\rangle\langle i+1|,\;\;\;i\in\mathbb{Z},\eeq
where $L$ and $R$ are the transition matrices, regarded as the sides of a quantum coin, satisfying the trace-preservation condition
\beq\label{bas2} L^*L+R^*R=I.\eeq
We say that the walk in (\ref{bas1}) is homogeneous, meaning that we have the same coin $(L,R)$ on every site, and the statistical meaning of the computation above is that the particle moves left with probability $\mathrm{Tr}(L\rho _i L^*)$ and moves right with probability $\mathrm{Tr}(R\rho _i R^*)$. If we extend this procedure by linearity we obtain an {\bf open quantum random walk (OQW)} on the integer line \cite{attal,attalhom}, see \cite{reviewOQW} for a recent survey. We have, implicit in such definition, a quantum trajectories formalism associated with a measurement of the position of the walk at every step. After the measurement, we normalize the result and repeat the process, making clear that we have a quantum evolution which is not unitary in general.

\medskip

The authors of \cite{attal} have formally defined OQWs and also stated a question on recurrence, which can be phrased as: given an OQW acting on some graph, how can we determine if a site is recurrent by looking at the transition matrices? At this point, one immediately needs to provide a proper definition of quantum recurrence and note that, in principle, more than one is available, see \cite{bbp,glv} for more on this. Generally speaking, recurrence problems are topics of interest in both open and closed (unitary) quantum dynamics and we seek a basic understanding of this kind of problem having in mind applications to quantum information and computation.

\medskip

A site recurrence criterion for  homogeneous OQWs on the line was established in the restricted case of coins given by order 2 diagonal or anti-diagonal matrices \cite{LS2015}, and later a condition is obtained in the case that $L$ and $R$ are normal matrices, see \cite{bbp,cgl}. In addition, the authors of \cite{bbp} provide a systematic study of first passage times of general OQWs, discussing in particular the irreducible case, giving further insight into the recurrence problem. More recently, in \cite{glv}, a discussion on quantum Markov chains (QMCs) and associated subspace recurrence problems is presented,   methods for determining site recurrence in terms of Schur functions are developed and, as OQWs are particular cases of QMCs, one obtains an useful understanding of the problem at hand from an analytic perspective. 

\medskip

Although the works mentioned above present a detailed study of the dynamics and statistics of OQWs, we believe one still misses, regarding homogeneous walks on the integers, a straightforward recurrence test, one which is somewhat comparable with the $p=1/2$ criterion obtained for the simple random walk. To the best of our knowledge, this kind of result  has been obtained only in special cases, as indicated above.

\medskip

In this work, we present a result which provides an answer for the question stated in \cite{attal}, namely, we describe a complete criterion for site recurrence of any homogeneous, nearest-neighbor OQW on the integer line for which the particle has an internal degree of dimension 2. We regard this to be the simplest, nonclassical version of the simple random walk. In addition, we describe a criterion regarding irreducible, homogeneous walks on the line for which the particle has an internal degree of arbitrary (finite) dimension, and also a result concerning walks for which the origin is an absorbing site. Such absorption problem regards the walk on the semi-infinite line induced by a coin: we inspect the probability that the walk will reach the origin eventually, by making use of a notion of monitored probability, see \cite{glv} and references therein for more on this notion. Due to the internal degree of freedom of the particle, one is able to see that recurrence and absorption may be related in a different manner than what is observed in the classical case.

\medskip

The structure of the proof relies in a crucial manner on a law of large numbers for OQWs \cite{attalhom}, see \cite{kummerer} for an earlier result, and on irreducibility properties of the associated operators \cite{bbp,carbone2, carbone1}. In particular, \cite{carbone2} contains some of the main ideas regarding the treatment of the reducible case for coins of dimension 2, which are revised and employed in this work.

\medskip

The contents of this work are organized as follows. In Section 2 we review OQWs, ergodicity and irreducibility properties. In Section 3 we discuss the notion of recurrence of OQWs adopted in the theory and we review some important facts. In Section 4 we prove an OQW version of Chung-Fuchs' Theorem, which states that a random walk on the line is recurrent if a law of large numbers condition holds. As an immediate consequence, we obtain a recurrence criterion for the irreducible case (Corollary \ref{critrecirredcoinrec}). Section 5 develops basic results leading to the main result of this work, Theorem \ref{teocrit}, which consists of a complete criterion for site recurrence of homogeneous OQWs on the integers with an internal degree of dimension 2. Section 6 discusses an absorption criterion for OQWs restricted to the semi-infinite line. Section 7 illustrates the results with examples.

\section{Preliminaries: OQWs, irreducibility and the auxiliary map}


Let $\mathcal{V}$ denote a complex, separable Hilbert space and let $\mathcal{B}(\mathcal{V})$ denote the bounded operators over $\mathcal{V}$. If $\dim\mathcal{V}=d<\infty$, we have that $\mathcal{B}(\mathcal{V})$ is identified with the order $d$ square matrices, denoted by $M_d(\mathbb{C})$. If $A\in M_d(\mathbb{C})$, let $A^*$ denote the conjugate transpose of $A$ and we write $A\geq 0$ to denote that $A$ is a positive semi-definite matrix, that is, $\langle Av,v\rangle\geq 0$, for all $v\in\mathbb{C}^d$. In this case we will simply say that $A$ is positive, and we write $A>0$ to denote strictly positive matrices. The space $\mathcal{B}(\mathcal{V})$ is the topological dual of its ideal $\mathcal{I}(\mathcal{V})$ of trace-class operators with trace norm
$$
\|\rho\|_1=\mathrm{Tr}(|\rho|),
\qquad 
|\rho|=\sqrt{\rho^*\rho},
$$
through the duality \cite[Lec. 6]{attal_lec}
$$
\langle \rho,X \rangle = \mathrm{Tr}(\rho X),
\qquad
\rho\in\mathcal{I}(\mathcal{V}),
\qquad 
X\in\mathcal{B}(\mathcal{V}).
$$
If we assume $\dim\mathcal{V}=d<\infty$, we have that $\mathcal{I}(\mathcal{V})=\mathcal{B}(\mathcal{V})=M_d(\mathbb{C})$. We refer the reader to \cite{attal_lec,glv} for more on the above preliminaries. 

\medskip

Moving towards the definition of OQW, the internal degree of freedom of some given particle, which we call  the {\bf coin space}, will be described by a complex Hilbert space $\mathcal{H}$, and in this work we will set $\mathcal{H}=\mathbb{C}^d$. We define the compact set of density operators over $\mathcal{H}$ by
$$\mathcal{D}(\mathcal{H})=\{\rho\in M_d(\mathbb{C}): \rho\geq 0, \mathrm{Tr}(\rho)=1\}.$$
In this work we will say {\bf integer line}, or simply the line, to refer to the set $\mathbb{Z}$ of integer numbers. Let $\mathcal{S}$ denote an auxiliary Hilbert space with an orthonormal basis $\{|i\rangle\}_{i\in \mathbb{Z}}$, which will be called the {\bf position space}. Define the following Banach subspace of $\mathcal{I}(\mathcal{H}\otimes\mathcal{S})$, given by
$$
\mathcal{I}_\mathcal{S}(\mathcal{H}) =
\{\rho\in\mathcal{I}(\mathcal{H}\otimes\mathcal{S}) : 
\textstyle \rho=\sum_{i\in \mathbb{Z}}\rho_i\otimes|i\rangle\langle i|\} =
\bigoplus_{i\in \mathbb{Z}} \mathcal{I}(\mathcal{H})\otimes|i\rangle\langle i|.
$$
For every $\rho\in\mathcal{I}_\mathcal{S}(\mathcal{H})$ we have that $\mathrm{Tr}(\rho)=\sum_{i\in \mathbb{Z}}\mathrm{Tr}(\rho_i)$. Then, define the following set of states,
$$
\mathcal{D}_\mathcal{S}(\mathcal{H}) = 
\{\rho\in\mathcal{I}_\mathcal{S}(\mathcal{H}):\rho\ge0,\;\mathrm{Tr}(\rho)=1\}.
$$
We call $\mathcal{D}_\mathcal{S}(\mathcal{H})$ the set of {\bf OQW densities}, and a density of the form 
\beq\label{bas000}
\eta\otimes |i\rangle\langle i|,\;\;\;\eta\geq 0,\;\;\;\mathrm{Tr}(\eta)=1,\eeq
corresponds to a particle located at site $i$, with internal degree $\eta$. An {\bf open quantum random walk (OQW)} acting on the line is the completely positive map given by
$$\Phi(\rho)=\sum_{i\in\mathbb{Z}}\Bigg(\sum_{j\in\mathbb{Z}} B_i^j\rho_j B_i^{j^*}\Bigg)\otimes|i\rangle\langle i|,\;\;\;\rho\in \mathcal{D}_\mathcal{S}(\mathcal{H}).$$
Above, matrix $B_i^j\in M_d(\mathbb{C})$ corresponds to the transition effect from site $j$ to $i$, and we assume that
$$\sum_i B_{i}^{j*}B_{i}^j=I,\;\;\;j\in\mathbb{Z},$$
which implies that $\mathrm{Tr}(\Phi(\rho))=\mathrm{Tr}(\rho)$, for every $\rho \in \mathcal{D}_\mathcal{S}(\mathcal{H})$, that is, $\Phi$ is trace-preserving. In \cite{attal} it is explained that if $\Phi$ is an OQW and $\rho\in\mathcal{D}_\mathcal{S}(\mathcal{H})$ then $\Phi(\rho)\in\mathcal{D}_\mathcal{S}(\mathcal{H})$. It is worth noting that if $d=1$, the OQW can be described by a stochastic matrix.

\medskip

In this work we will focus on homogeneous walks: the particle moves exactly one position to the left, or to the right, at each given step, and the coin is constant, i.e., the same transition rule is applied at each site. This allows us to write
\beq\label{oqwh1}\Phi(\rho)=\sum_{i\in\mathbb{Z}}\Big(R\rho_{i-1} R^* + L\rho_{i+1} L^* \Big)\otimes |i\rangle\langle i|,\;\;\;\rho\in \mathcal{D}_\mathcal{S}(\mathcal{H}),\eeq
which is the formal definition of the walk described in the Introduction via eq. (\ref{bas1}). We say that the pair of matrices $(L,R)$ is a \textbf{coin} if eq. (\ref{bas2}) holds. If $\mathcal{H}=\mathbb{C}^d$ for some $d$, the \textbf{dimension} of the coin is defined to be $d$. 

\medskip

{\bf Definition.} The {\bf OQW induced by a coin $(L,R)$} is the homogeneous, nearest-neighbor OQW on the integer line with left and right transitions given by $L$ and $R$, respectively. The completely positive, trace-preserving map describing the walk is given by (\ref{oqwh1}). We also say that the coin \textbf{spans} the OQW, see Fig. 1.

\begin{center}
\begin{figure}[ht]\label{fig1}
\begin{tikzpicture}
[->,>=stealth',shorten >=1pt,auto,node distance=2.0cm,
                    semithick]
    \node[state,draw=none] (1) {$\cdots$};
    \node[main node] (2) [ right = 2.0cm and 2.0cm of 1]  {$i$};
    \node[main node] (3) [ right = 2.0cm and 2.0cm of 2] {$i+1$};
\node[state,draw=none] (d1) [right=2.0cm and 2.0cm of 3] {$\cdots$};

    \path[draw,thick]
    (1) edge   [bend left]                     node {$R\cdot R^*$} (2)
    (2) edge   [bend left]      node {$L\cdot L^*$} (1)
    (2) edge   [bend left]                     node {$R\cdot R^*$} (3)
    (3) edge   [bend left]      node {$L\cdot L^*$} (2)
        (3) edge   [bend left]     node {$R\cdot R^*$} (d1)
        (d1) edge   [bend left]     node {$L\cdot L^*$} (3)

    ;

\end{tikzpicture}
\tcaption{A homogeneous, nearest-neighbor OQW on the line, where matrices $L$, $R$ are such that $L^*L+R^*R=I$. If at time $n$ we are at site $i$ with initial density $\rho_i\otimes |i\rangle\langle i|$, then at time $n+1$, either we move to site $i-1$ with probability $\mathrm{Tr}(L\rho_i L^*)$ and the new density becomes, after normalization, $L\rho_i L^*/\mathrm{Tr}(L\rho_i L^*)\otimes |i-1\rangle\langle i-1|$, or we move to site $i+1$ with probability $\mathrm{Tr}(R\rho_i R^*)$ and the new density becomes $R\rho_i R^*/\mathrm{Tr}(R\rho_i R^*)\otimes |i+1\rangle\langle i+1|$.}
\end{figure}
\end{center}

A \textbf{quantum trajectory} of a homogeneous OQW on the line, spanned by a coin $(L,R)$ and starting from a state $\rho$ of the form $\sum_{i \in \mathbb{Z}}\rho_i \otimes |i\rangle\langle i|,$
is any path generated by the Markov chain $(x_n,\rho_n)_{n\geq 0}$, where $x_n$ denotes the position of the particle at time $n$ and $\rho_n$ its internal degree. The transition probabilities are given by
$$
\mathbb{P}\left(\, ( x_{n+1},\rho_{n+1}) = \left(i+1, \frac{R\sigma R^*}{\mathrm{Tr}(R\sigma R^*)}\right) \; \Bigg| \; (x_n,\rho_n) = (i, \sigma)\,\right) = \mathrm{Tr}(R\sigma R^*),
$$
$$
\mathbb{P}\left(\, (x_{n+1},\rho_{n+1}) = \left(i-1, \frac{L\sigma L^*}{\mathrm{Tr}(L\sigma L^*)}\right) \; \Bigg| \; (x_n,\rho_n) = (i, \sigma)\,\right) = \mathrm{Tr}(L\sigma L^*),
$$
for every $i \in \mathbb{Z}$, $\sigma \in \mathcal{D}(\mathcal{H})$, and initial law
$$
\mathbb{P}\left(\, (x_{0},\rho_{0}) = \left(i, \frac{\rho_i}{Tr\,\rho_i}\right) \; \right) = \mathrm{Tr}\,\rho_i,
$$
and all other transition probabilities being equal to $0$. At time $0$ we will usually consider a particle located at some site $i$, as in (\ref{bas000}). It is worth noting that while the pair $(x_n,\rho_n)_{n\geq 0}$ is in fact a Markov chain in the classical sense, the process given by the position alone $(x_n)_{n\geq 0}$, with a fixed initial density $\rho$, is not. As a consequence, many results of the theory of OQWs regarding site recurrence are not immediately implied by the classical theory, see more on this in Section 3.

\medskip

A \textbf{pure density} in $\mathcal{H}$ is a density $\rho$ that is a projection of rank 1, i.e., there exists $|\varphi\rangle\in\mathbb{C}^d$ such that $\rho = |\varphi\rangle\langle\varphi|$. A \textbf{faithful density} in $\mathcal{H}$ is a density which is strictly positive.  Note in particular that for a density $\rho$ of dimension $2$, $\rho$ is faithful if, and only if, it is a nontrivial convex sum of two pure densities. 

\medskip

Now we recall the important notion of irreducibility, following \cite{carbone2,carbone1}. Let $T$ denote a positive map (that is, such that if $X\geq 0$ then $T(X)\geq 0$) acting on the space $\mathcal{I}(\mathcal{V})$ of trace-class operators of a Hilbert space $\mathcal{V}$. We say that $T$ is {\bf irreducible} if the only orthogonal projections $P$ such that $T(P \mathcal{I}(\mathcal{V})P)\subset P \mathcal{I}(\mathcal{V})P$, are $P=0$ and $P=I$. There are several equivalent definitions of irreducibility, but for our purposes we will restrict to comment only on aspects which are needed in this work. Regarding homogeneous OQWs induced by order 2 coins, the following proposition allows us to determine if a walk is irreducible:

\begin{pro}[\cite{carbone2}, Prop. 6.12]
Let $(L,R)$ be a coin of dimension 2.
Define
$$
W = \{\text{common eigenvectors of } LR \text{ and } RL\}.
$$
The OQW induced by $(L,R)$ is reducible if, and only if, one of the following facts holds:
\begin{enumerate}
\item
$W$ contains an eigenvector of $L$ or $R$.
\item 
$W= \mathbb{C}u \bigcup \mathbb{C}v \backslash \{0\}$, for some linearly independent vectors $u$ and $v$ satisfying
$Lu, Ru \in \mathbb{C}v$ and $Lv, Rv \in \mathbb{C}u$.
\end{enumerate}

\end{pro}

The following are two important results on irreducibility:
\begin{theorem}\label{cp_irr} Let $T$ be a completely positive map acting on the trace-class operators of a Hilbert space.
\begin{itemize} 
\item[a)] \cite[ Rk. 3.6]{carbone2} If $T$ admits a unique invariant state and such state is faithful, then $T$ is irreducible.
\item[b)] \cite[Thm. 3.14]{carbone1} If $T$ is irreducible and has an invariant state, then it is unique and faithful. 
\end{itemize}
\end{theorem}

Given an OQW $\Phi$ spanned by a coin $(L,R)$, we define the \textbf{auxiliary map} $\mathfrak{L}_\Phi$ by
$$\begin{array}{l}
\mathfrak{L}_\Phi :  M_d(\mathbb{C}) \to M_d(\mathbb{C}),\;\;\;
\mathfrak{L}_\Phi(\rho) =  L\rho L^* +R \rho R^*.
\end{array}
$$
We will usually write $\mathfrak{L}_\Phi=\mathfrak{L}$ if the associated OQW $\Phi$ is clear from context. Let 
$$\mathcal{P}_l(0)=\Big\{\pi=(s_1,\dots,s_l)\in\{-1,1\}^l:\sum_{i=1}^l s_i=0\Big\},\;\;\;\mathcal{P}(0)=\bigcup_{l\in\mathbb{N}}\mathcal{P}_l(0)$$
so that if $B_{1}=R$ and $B_{-1}=L$, the notation
$$B_\pi=B_{s_l}\cdots B_{s_1},\;\;\;\pi=(s_1,\dots,s_l)\in\mathcal{P}_l(0)$$
describes a matrix product such that the number of $L$'s is equal to the number of $R$'s.  We recall the following:

\begin{pro}\label{corCarb} Let $(L,R)$ be a coin of finite dimension that spans an OQW $\Phi$, and let $\mathfrak{L}$ be its auxiliary map. Then:
\begin{itemize} 
\item[i)] \cite[Prop. 4.1]{carbone2}
a) $\mathfrak{L}$ is irreducible if, and only if, $L$ and $R$   
have no invariant closed subspace in common, apart from $\{0\}$ and $\mathcal{H}$.

b) $\Phi$ is irreducible if, and only if, the set of matrix products $\{B_{\pi_0}: \pi_0\in\mathcal{P}(0)\}$ have no invariant closed subspace in common, apart from $\{0\}$ and $\mathcal{H}$. 

\item[ii)] \cite[Cor. 4.2]{carbone2} If $\Phi$ is irreducible then $\mathfrak{L}$ is irreducible.
\end{itemize}
\end{pro}
The auxiliary map is an essential object in the study of recurrence, as it will be illustrated later in this work.

\section{Recurrence for OQWs}\label{sec2a}

For any site $|i\rangle$, $\rho\in\mathcal{D}_\mathcal{S}(\mathcal{H})$, let ${P}_i(\rho)=(I_d\otimes |i\rangle\langle i|)\rho(I_d\otimes |i\rangle\langle i|)$, where $I_d$ denotes the order $d$ identity matrix. Given an OQW $\Phi$ and associated quantum trajectories $(x_n,\rho_n)_{n\geq 0}$, the {\bf probability of reaching site $|j\rangle$}  at the $n$-th step, given that the walk started at site $|i\rangle$ with initial density $\rho$, is given by
$$
\mathbb{P}_{i,\rho}(x_n=j)=\mathrm{Tr}\Big({P}_j\Phi^n(\rho\otimes|i\rangle\langle i|)\Big),\;\;\;\rho\in\mathcal{D}(\mathcal{H})
$$

\medskip

{\bf Definition.} We say that an OQW is: 
\begin{itemize}
\item
 {\bf recurrent at site  $i$ with initial density $\rho$}  if

\beq\label{defr1}
\sum_{n=0}^{\infty} \mathbb{P}_{i,\rho}(x_{n}=i) = \infty.
\eeq
\item
 {\bf transient at site  $i$ with initial density $\rho$} if

\beq
\sum_{n=0}^{\infty} \mathbb{P}_{i,\rho}(x_{n}=i) < \infty.
\eeq
\end{itemize}
We also say that the walk is {\bf site-recurrent} and  {\bf site-transient}, respectively, with respect to $i$ and density $\rho$. We say that an OQW is:
\begin{itemize}
\item ${\bf recurrent}$ if it is recurrent at every site $i$, with respect to every initial density,

\item ${\bf transient}$ if it is  transient at every site $i$, with respect to every initial density. 
\end{itemize}
Finally, regarding the homogeneous case, we say that {\bf a coin $(B,C)$ is recurrent} if the OQW induced by such coin is recurrent, and we say that {\bf a coin $(B,C)$ is transient} if the OQW induced by such coin is transient.

\medskip

It is a simple matter to show that there are OQWs which are neither recurrent nor transient. However, regarding irreducible OQWs, we have the following important dichotomy:
\begin{pro}\label{dichot}[\cite{bbp}, Cor. 3.10]
Every irreducible OQW is either transient or recurrent.
\end{pro}

If denote by $n_i= \#\{n \geq 1 | x_n = i\}$, the number of visits to $i$, and by $\mathbb{E}$ the expected value of a random variable, then
$$
\sum_{n=0}^\infty \mathbb{P}_{i,\sigma}(x_n = i) = 
\sum_{n=0}^\infty \mathbb{E}_{i,\sigma}(1_{\{x_n = i\}}) = 
\mathbb{E}_{i,\sigma}\sum_{n=0}^\infty 1_{\{x_n = i\}} = 
\mathbb{E}_{i,\sigma}( n_i),
$$
which makes clear that the present notion of recurrence is given in terms of the  mean number of visits to a site.

\medskip

We can also study recurrence in terms of generating functions as follows (a formal discussion on the analytic behavior of such series can be seen in \cite{glv}). If we define 
$${s}_{ii}(z)=\sum_{n\geq 0}{P}_i(z\Phi)^n{P}_i={P}_i(I-z\Phi)^{-1}{P}_i,\;\;\;|z|<1,$$
we can write (\ref{defr1}) as
$$\lim_{z\uparrow 1}\mathrm{Tr}({s}_{ii}(z)\rho) = \infty.$$
In this work, whenever we refer to recurrence, we mean the notion given by expression (\ref{defr1}), unless otherwise noted. This notion is sometimes called  {\bf recurrence in the sense of P\'olya}.

\medskip

Also in terms of generating functions, the following notion will be used for the study of absorption in the setting of OQWs. For any site $|i\rangle$, $\rho\in\mathcal{D}_\mathcal{S}(\mathcal{H})$, let  ${Q}_i=[I_d\otimes(\mathbb{I}-|i\rangle\langle i|)]\rho[I_d\otimes(\mathbb{I}-|i\rangle\langle i|)]$, where $\mathbb{I}$ denotes the identity on the position space $\mathcal{S}$. That is, $Q_i$ discards any information contained on site $i$. The {\bf probability of reaching site $|j\rangle$  for the first time} at the $n$-th step, given that the walk started at site $|i\rangle$ with initial density $\rho$ is given by
\beq\label{fprob1}
\mathbb{F}_{i,\rho}(x_n=j)=\mathrm{Tr}\Big({P}_j\Phi({Q}_j\Phi)^{n-1}(\rho\otimes |i\rangle\langle i|)\Big),\;\;\;\rho\in\mathcal{D}(\mathcal{H}),
\eeq
and if we define
$$f_{ii}(z)=\sum_{n\geq 0}{P}_i\Phi(z{Q}_i\Phi)^n{P}_i={P}_i\Phi(1-z{Q}_i\Phi)^{-1}{P}_i,\;\;\;|z|<1,$$
the probability that a walk ever returns to site $i$, with initial density $\rho$, can be written as
$$\mathbb{P}_{i,\rho}(t_i<\infty)=\sum_{n\geq 1}\mathbb{F}_{i,\rho}(x_{n}=i)=\lim_{z\uparrow 1}\mathrm{Tr}(f_{ii}(z)\rho),\;\;\;t_i=\min\{n>0: x_n=i\},$$
and we say that site $i$ is {\bf monitored-recurrent} with respect to $\rho$ if $\mathbb{P}_{i,\rho}(t_i<\infty)=1$. At this point, a natural question regards the relation between recurrence in terms of (\ref{defr1}) and monitored recurrence in the setting of OQWs. Let us briefly discuss this in what follows.

\subsection{On the probability of first return and mean number of visits}\label{sbsss}

It is worth recalling that in the classical case of Markov chains we have the renewal equation \cite{grimmett}
$$P_{ii}(z)=\frac{1}{1-F_{ii}(z)},$$
where $P_{ii}$ and $F_{ii}$ are the classical generating functions associated with visits to $i$, and first returns to $i$, respectively. The above relation  implies that
$$\lim_{z\uparrow 1}F_{ii}(z)=1 \;\Longleftrightarrow\; \lim_{z\uparrow 1}P_{ii}(z)=\infty.$$
In words: in the classical setting, to say that the return probability to site $i$ is equal to 1 is equivalent to say that the mean number of visits to $i$ is infinite. On the other hand, in the setting of OQWs, the inspection of an analogous correspondence needs to take in consideration the internal degrees of freedom of the sites. More precisely, it is instructive to recall the following items: 
\begin{enumerate}

\item $\mathbb{P}_{i,\rho}(t_i<\infty)=1$ for every $\rho$ implies that $\mathbb{E}_{i,\rho}(n_i)=\infty$, for every $\rho$ [\cite{cgl}, Prop. 6.1]. 
\item $\mathbb{P}_{i,\rho}(t_i<\infty)=1$ for some faithful $\rho$ implies that $\mathbb{E}_{i,\eta}(n_i)=\infty$ for every density $\eta$, see [\cite{bbp}, Cor. 3.5].
\item $\mathbb{P}_{i,\rho}(t_j<\infty)=1$ for some faithful $\rho$ implies that $\mathbb{P}_{i,\eta}(t_j<\infty)=1$ for every density $\eta$, see [\cite{bbp}, Cor. 3.5].
\item $\mathbb{E}_{i,\rho}(n_i)=\infty$ for some $\rho$ implies $\mathbb{P}_{i,\rho'}(t_i<\infty)=1$, for some $\rho'$ which is accessible from $\rho$. See [\cite{cgl}, Prop. 6.2] and [\cite{bbp}, Rk. 3.8 and Ex. 5.1]. 
We recall that $\rho'$ is accessible from $\rho$ if there is a path, described by a product of transition matrices, e.g. $B=LLRRLR$, such that $\rho'=B\rho B^*/Tr(B\rho B^*)$.
\item Supposing the OQW is irreducible, if for some density $\rho$ we have 
$\mathbb{E}_{i,\rho}(n_i)=\infty$ then this is true for every density and $\mathbb{P}_{i,\rho}(t_i<\infty)=1$ for every $\rho$ [\cite{bbp}, Teo 3.1].
\end{enumerate}

In particular, monitored recurrence with respect to every density (which is the main definition adopted in \cite{cgl,LS2015}) implies recurrence with respect to every density, in the sense of this work. With respect to the above notions, we prove the following result, further simplifying the relation between the mean number of visits to a vertex and the first return probability in the case of irreducible, homogeneous walks:

\begin{pro}\label{equivhom} For any irreducible, homogeneous OQW induced by a coin $(L,R)$, we are in one of the following situations:
\begin{enumerate}
\item For any $i$, $j$ sites and every density $\rho$, we have $\mathbb{E}_{i,\rho}(n_j)=\infty$ and $\mathbb{P}_{i,\rho}(t_j<\infty)=1$,
\item For any $i$, $j$ sites and every density $\rho$, we have $\mathbb{E}_{i,\rho}(n_j)<\infty$ and $\mathbb{P}_{i,\rho}(t_j<\infty)<1$.
\end{enumerate}
\end{pro}
{\bf Proof.} The two cases stated above are described in [\cite{bbp}, Thm. 3.1], along with the following situation: $\mathbb{E}_{i,\rho}(n_j)<\infty$ for any $i$, $j$ sites, $\rho$ density, and there are $\rho, \rho'$ densities, $\rho$ non-faithful, such that for some site $i$, we have  $\mathbb{P}_{i,\rho}(t_i<\infty)=1$ and $\mathbb{P}_{i,\rho'}(t_i<\infty)<1$. The proposition will be proved by showing that, under the hypothesis, this latter case is impossible.

\medskip

The idea of the proof is to show that the assumption $\mathbb{P}_{i,\rho}(t_i<\infty)=1$ implies that there is $\eta$ faithful located at any  $j\neq i$ such that $\mathbb{P}_{j,\eta}(t_i<\infty)=1$. By item (3) preceeding this proposition, this probability must be equal to 1 for every density. Then $\mathbb{P}_{i,\eta'}(t_j<\infty)=1$ for some $\eta'$ faithful (and therefore for every density) and this implies that $\mathbb{P}_{j,\eta}(t_j<\infty)=1$ for every density $\eta$. By item (1) stated above, we conclude $\mathbb{E}_{i,\eta}(n_i)=\infty$, a contradiction.

\medskip

As the walk is homogeneous, we may assume $i=0$. Note that, after one step of the walk, we have
$$1=\mathbb{P}_{-1,\frac{L\rho L^*}{\mathrm{Tr}(L\rho L^*)}}(t_0<\infty)\cdot\mathrm{Tr}(L\rho L^*)+
       \mathbb{P}_{1,\frac{R\rho R^*}{\mathrm{Tr}(R\rho R^*)}}(t_0<\infty)\cdot\mathrm{Tr}(R\rho R^*).$$
By the trace preservation condition, we must have that the probabilities on the right are equal to 1. After 2 steps of the walk, we have
$$
1=\mathrm{Tr}(LR\rho R^*L^*)+ \mathrm{Tr}(RL\rho L^*R^*)$$
$$+\mathbb{P}_{-2,\frac{L^2\rho L^{2*}}{\mathrm{Tr}(L^2\rho L^{2*})}}(t_0<\infty)\cdot \mathrm{Tr}(L^2\rho L^{2*})
+
\mathbb{P}_{2,\frac{R^2\rho R^{2*}}{\mathrm{Tr}(R^2\rho R^{2*})}}(t_0<\infty)\cdot \mathrm{Tr}(R^2\rho R^{2*}).$$
As in the first step, due to trace preservation, we have that both $\mathbb{P}_{-2}$ and $\mathbb{P}_{2}$ must be equal to 1. We may repeat the above reasoning so that for every $j\neq 0$ we can obtain, after a sufficient number of steps, a set of densities $\eta_1, \eta_2,\dots,\eta_{k}$ such that $\mathbb{P}_{j,\eta_m}(t_0<\infty)=1$, $m=1,\dots,k$, and from which we can obtain a faithful density $\alpha$ located at $j$ via a convex combination. This is always possible, due to the irreducibility assumption. Therefore, we have $\mathbb{P}_{j,\alpha}(t_0<\infty)=1$ and we are done.

\qed

\begin{remark} We note that site recurrence refers to the probability that, given an initial density concentrated on site $i$, say, $\eta\otimes |i\rangle\langle i|$, the walk will eventually return to site $i$, landing on such site with {\bf any} density matrix. This is in contrast to the {\bf state recurrence} problem, which consists of determining the probability that the walk will return to $i$ with the same initial density $\eta$, see \cite{glv} for more on this.\end{remark}

\section{OQW version of Chung-Fuchs' Theorem and the irreducible case}\label{sec3}

In this section we consider some results over the Markov chain $(x_n,\rho_n)_{n\geq 0}$ induced by the quantum trajectories of a homogeneous OQW on the line. This will lead us to a site-recurrence criterion for the irreducible case. Lemma \ref{lemChung}, together with Theorem \ref{CFteoOWQ}, will allow us to make practical use of the law of large numbers proved in \cite{attalhom}, see \cite{kummerer} for an earlier result. The proofs below are inspired by the classical result \cite{chungf}, also see \cite{durrett}. 

\medskip

\begin{lemma}\label{lemChung} Let $(x_n,\rho_n)_{n\geq 0}$ denote the trajectories of a homogeneous OQW on the integer line. Let $\varepsilon>0$ and $m>1$ be an integer. Then, for every $\rho \in \mathcal{D}=\mathcal{D}(\mathcal{H})$,
$$
\sum_{n=0}^\infty \mathbb{P}_{0,\rho}(|x_n|<m\varepsilon) 
\leq
 2m\cdot \max_{\sigma \in \mathcal{D}}\sum_{n=0}^\infty \mathbb{P}_{0,\sigma}(|x_n|<\varepsilon).
$$
\end{lemma} 

\textbf{Proof.} As $ (-m,m) \subset \bigcup_{k=-m}^{m-1}[k,k+1) $, we have
\beq\label{eqlemchung}
\sum_{n=0}^\infty \mathbb{P}_{0,\rho}(|x_n|<m\varepsilon) \leq 
\sum_{n=0}^\infty \sum_{k=-m}^{m-1}\mathbb{P}_{0,\rho}(k\varepsilon \leq x_n < (k+1)\varepsilon).
\eeq

Let $T_k = \inf \{ l \geq 0 : k\varepsilon \leq x_l < (k+1)\varepsilon \}$ and 
$\Sigma(\rho,k,\varepsilon) = \sum_{n=0}^\infty \mathbb{P}_{0,\rho}(k\varepsilon \leq x_n < (k+1)\varepsilon)$. Then, by summing over the value of $T_k$,
$$
\Sigma(\rho,k,\varepsilon)
=
\sum_{n=0}^\infty \sum_{l=0}^\infty \mathbb{P}_{0,\rho}(k\varepsilon \leq x_n < (k+1)\varepsilon,T_k=l).
$$
If $T_k = l, $ then $k\varepsilon \leq x_l < (k+1)\varepsilon$, 
which implies $- (k+1)\varepsilon< -x_l \leq -k\varepsilon $
 and $- \varepsilon< x_n -x_l < \varepsilon $. Consequently, by Fubini's Theorem,
$$
\Sigma(\rho,k,\varepsilon)
=
\sum_{n=0}^\infty \sum_{l=0}^\infty \mathbb{P}_{0,\rho}(|x_n - x_l |<\varepsilon,T_k=l)=\sum_{l=0}^\infty \sum_{n=l}^\infty \mathbb{P}_{0,\rho}(|x_n - x_l |<\varepsilon,T_k=l).
$$

As the events $\{T_k=l\}$ and$\{|x_n-x_l|<\varepsilon \}$ are independent, we have
$$
\Sigma(\rho,k,\varepsilon)
=
\sum_{l=0}^\infty \mathbb{P}_{0,\rho}(T_k=l)
 \sum_{n=l}^\infty \mathbb{P}_{0,\rho}(|x_n - x_l |<\varepsilon).
$$	
By summing according to the value of $(x_l,\rho_l)$ and applying Fubini's theorem,
$$
\Sigma(\rho,k,\varepsilon)
=
\sum_{l=0}^\infty \mathbb{P}_{0,\rho}(T_k=l)
 \sum_{n=l}^\infty 
\sum_{\sigma\in\mathcal{D}} \sum_{j \in \mathbb{Z}}
\mathbb{P}_{0,\rho}(|x_n - x_l |<\varepsilon \,, \, (x_l,\rho_l) =( j,\sigma))
$$
$$
=
\sum_{l=0}^\infty \mathbb{P}_{0,\rho}(T_k=l)
\sum_{\sigma\in\mathcal{D}} \sum_{j \in \mathbb{Z}} \sum_{n=l}^\infty 
\mathbb{P}_{0,\rho}(|x_n - x_l |<\varepsilon| (x_l,\rho_l) =( j,\sigma))
\cdot \mathbb{P}_{0,\rho}( (x_l, \rho_l) =( j,\sigma)).
$$
Note that the summation over $\sigma\in\mathcal{D}$ above makes sense, as $\mathbb{P}_{0,\rho}( (x_l,\rho_l) =( j,\sigma))$ is supported on a finite set as a function of $\sigma$ (i.e., just a finite set of densities are possible via the trajectories of the walk). As the walk is homogeneous in space, we have
$$
\Sigma(\rho,k,\varepsilon)
=
\sum_{l=0}^\infty \mathbb{P}_{0,\rho}(T_k=l)
\sum_{\sigma\in\mathcal{D}} \sum_{j \in \mathbb{Z}} \sum_{n=l}^\infty 
\mathbb{P}_{j,\sigma}(|x_{n-l} - x_0 |<\varepsilon)
\cdot \mathbb{P}_{0,\rho}( (x_l,\rho_l) =( j,\sigma)).
$$
By letting $s=n-l$, we have
$$
\Sigma(\rho,k,\varepsilon)
=
\sum_{l=0}^\infty \mathbb{P}_{0,\rho}(T_k=l)
\sum_{\sigma\in\mathcal{D}} \sum_{j \in \mathbb{Z}}
\sum_{s=0}^\infty 
\mathbb{P}_{j,\sigma}(|x_{s} -j |<\varepsilon)
\cdot
 \mathbb{P}_{0,\rho}( (x_l,\rho_l) =( j,\sigma))
$$
$$
=
\sum_{l=0}^\infty \mathbb{P}_{0,\rho}(T_k=l)
\sum_{\sigma\in\mathcal{D}}
\sum_{s=0}^\infty 
\mathbb{P}_{0,\sigma}(|x_{s} |<\varepsilon)
\cdot
 \sum_{j \in \mathbb{Z}}
 \mathbb{P}_{0,\rho}( (x_l,\rho_l) =( j,\sigma)).
$$
Therefore,
$$
\Sigma(\rho,k,\varepsilon)
\leq
\sum_{l=0}^\infty \mathbb{P}_{0,\rho}(T_k=l)
\max_{\alpha \in \mathcal{D}}
\sum_{s=0}^\infty 
\mathbb{P}_{0,\alpha}(|x_{s} |<\varepsilon)
\cdot
\sum_{\sigma\in\mathcal{D}}
 \sum_{j \in \mathbb{Z}}
 \mathbb{P}_{0,\rho}( (x_l,\rho_l) =( j,\sigma))
$$
$$
=
\sum_{l=0}^\infty \mathbb{P}_{0,\rho}(T_k=l)
\max_{\alpha \in \mathcal{D}}
\sum_{s=0}^\infty 
\mathbb{P}_{0,\alpha}(|x_{s} |<\varepsilon)
= \max_{\alpha \in \mathcal{D}}
\sum_{s=0}^\infty 
\mathbb{P}_{0,\alpha}(|x_{s} |<\varepsilon).
$$

Finally, by the above inequality and (\ref{eqlemchung}), we have
$$
\sum_{n=0}^\infty \mathbb{P}_{0,\rho}(|x_n|<m\varepsilon) \leq 
 \sum_{k=-m}^{m-1}
\Sigma(\rho,k,\varepsilon)
\leq
 \sum_{k=-m}^{m-1}
\max_{\alpha \in \mathcal{D}}
\sum_{s=0}^\infty 
\mathbb{P}_{0,\alpha}(|x_{s} |<\varepsilon)
=
2m \cdot
\max_{\alpha \in \mathcal{D}}
\sum_{s=0}^\infty 
\mathbb{P}_{0,\alpha}(|x_{s} |<\varepsilon).
$$

\cqfd

\begin{theorem}[Chung-Fuchs Theorem for OQWs]\label{CFteoOWQ}
Given an OQW on the integer line with quantum trajectories $(x_n,\rho_n)_{n\geq 0}$, if $x_n / n \rightarrow 0 $ in probability, then there exists $\sigma \in \mathcal{D}$ such that we have recurrence at site $0$ with initial density $\sigma$, i.e.,  $\sum_{n=0}^\infty \mathbb{P}_{0,\sigma}(x_{n}=0) = \infty. $ 
\end{theorem}

\textbf{Proof.} 
Let $u_n^\rho(x) = \mathbb{P}_{0,\rho}(|x_{n} |<x).$
 Let $\sigma \in \mathcal{D}$ such that 
$$
\sum_{n=0}^\infty u_n^\sigma (1)
=\max_{\alpha \in \mathcal{D}} \sum_{n=0}^\infty u_n^\alpha (1).
$$
Let $m>1$ be an integer and let $\rho \in \mathcal{D}$. By Lemma \ref{lemChung} we have
$$
2m \cdot\sum_{n=0}^\infty u_n^\sigma (1)
=
2m \cdot \max_{\alpha \in \mathcal{D}} \sum_{n=0}^\infty u_n^\alpha (1)
\geq
\sum_{n=0}^\infty u_n^\rho (m\cdot 1).
$$
Let $A>0$ be an integer. Then
\beq\label{eqteoChung}
 \sum_{n=0}^\infty \mathbb{P}_{0,\sigma}(x_{n}=0) =  \sum_{n=0}^\infty u_n^\sigma (1) 
\geq
\frac{1}{2m}\sum_{n=0}^\infty u_n^\rho (m)
\geq
\frac{1}{2m}\sum_{n=0}^{Am} u_n^\rho (m)
\geq
\frac{1}{2m}\sum_{n=0}^{Am} u_n^\rho (n/A),
\eeq
since $u_n^\rho (x) \geq 0$ and is increasing in $x$. Now note that $
 u_n^\rho (n/A) = \mathbb{P}_{0,\rho}(|x_{n} |<n/A)=\mathbb{P}_{0,\rho}(|x_{n} |/n<1/A)
$ so, by the weak law hypothesis, we have 
$$
\lim_{n\to \infty} u_n^\rho (n/A) =1,
$$
which implies
$$
\lim_{m\to \infty}\frac{1}{2m}\sum_{n=0}^{Am} u_n^\rho (n/A) 
= \frac{A}{2}\cdot\lim_{m\to \infty}\frac{\sum_{n=0}^{Am} u_n^\rho (n/A)}{Am}
=\frac{A}{2}.
$$
By eq. (\ref{eqteoChung}), we have
$$
 \sum_{n=0}^\infty \mathbb{P}_{0,\sigma}(x_{n}=0) \geq \frac{A}{2}
$$
and, as $A$ is arbitrary, we must have $ \sum_{n=0}^\infty \mathbb{P}_{0,\sigma}(x_{n}=0) = \infty$.

\cqfd

\begin{cor}\label{critrecirredcoinrec}
Let $(L,R)$ be a coin inducing a homogeneous, irreducible OQW on the line. Let  $\rho_{\infty}$ be the unique invariant state of  the auxiliary map. Then,
$$
 \mathrm{Tr}(L^*L\rho_{\infty}) = \frac{1}{2}  \Rightarrow  (L,R) \text{ is recurrent.}
$$
The same implication holds with $R$ instead of $L$ since $\mathrm{Tr}(R^*R\rho)=\frac{1}{2}\;\Leftrightarrow\;\mathrm{Tr}(L^*L\rho)=\frac{1}{2}$.
\end{cor}
\textbf{Proof.}   
As the induced walk is irreducible,  by Proposition \ref{corCarb} the auxiliary map $\mathfrak{L}$ also  is irreducible, so it admits a unique invariant state $\rho_{\infty}$.  Let $(x_n,\rho_n)_{n\geq 0}$ be the Markov chain that generates the quantum trajectories of the OQW spanned by coin $(L,R)$.
 Then by [\cite{attalhom}, Thm. 5.2], we have that the law of large numbers $\lim_{n \to \infty} \frac{x_n}{n} =\mu$ holds almost surely with $|\mu| = |1 -2 \mathrm{Tr}(L^*L\rho_{\infty})|=0$. Then, convergence in probability holds, so the hypothesis of Theorem \ref{CFteoOWQ} is satisfied and we have recurrence at site $0$ for some initial density $\sigma$. As the OQW is irreducible, it is recurrent  by Proposition \ref{dichot}.
\cqfd


\medskip

The case for which $\mathrm{Tr}(L^*L\rho_{\infty}) \neq \frac{1}{2}$ leads to transience as long as the auxiliary map of the walk admits a unique invariant state:
\begin{pro}\label{critrecirredcointrans}
Let $(L,R)$ be a coin inducing a homogeneous OQW $\Phi$ on the line. Assume that the auxiliary map admits a unique invariant state $\rho_{\infty}$. Then,
$$
 \mathrm{Tr}(L^*L\rho_{\infty}) \neq \frac{1}{2}  \Rightarrow  (L,R) \text{ is transient.}
$$
\end{pro}
\textbf{Proof.}   
 Let $(x_n,\rho_n)_{n\geq 0}$ be the Markov chain that generates the quantum trajectories of the OQW spanned by coin $(L,R)$. Then by [\cite{attalhom}, Thm. 5.2], we have that the law of large numbers $\lim_{n \to \infty} \frac{x_n}{n} =\mu$ holds almost surely with $|\mu| = |1 -2 \mathrm{Tr}(L^*L\rho_{\infty})|\neq 0$. Then  $|x_n|  \rightarrow \infty$ $\mathbb{P}_{x_0,\rho_0}$-a.s., implying that  $\mathbb{P}_{0,\rho}(x_n = 0 \text{ i.o.})=0$ for any $\rho$. Therefore, the mean number of visits to $0$ is finite, that is, $(L,R)$ is transient.
\cqfd

\section{Recurrence criteria for order 2 coins}\label{sec4}

The following preliminary results will lead us to a complete set of recurrence criteria for coins in $M_2(\mathbb{C})$. This is stated in Theorem \ref{teocrit}.

\begin{lemma}\label{lemrecfaithful}
Regarding an OQW induced by a coin $(L,R)$ of dimension 2, we have:
 \begin{enumerate}
 \item Recurrence w.r.t a pure density implies recurrence w.r.t. all faithful densities.
\item  Recurrence w.r.t a non-pure density implies recurrence w.r.t at least one pure density.
\end{enumerate}
\end{lemma}
\textbf{Proof.} (1) Let  $|a\rangle$ be a  unit vector such that $(L,R)$  is recurrent with respect to $|a\rangle\langle a|.$
 Let $ |b\rangle$ such that $\{|a\rangle, |b\rangle\}$ is an orthonormal basis of $\mathbb{C}^2$ and let $\rho $  be a faithful density, which can be written in terms of such basis as  $\rho = \begin{bmatrix} \rho_{11} & \rho_{12} \\ \overline{\rho_{12}} & 1-\rho_{11}\end{bmatrix}$. Then, $0<\rho_{11}<1$ and $\rho_{11}(1-\rho_{11})-|\rho_{12}|^2>0$. Let $\varepsilon$ such that $0<\varepsilon <\rho_{11}-\frac{|\rho_{12}|^2}{1-\rho_{11}}$. Then $\rho_\varepsilon =\rho - \varepsilon \cdot|a\rangle\langle a| $ is  positive-definite and we have
$$
\mathbb{P}_{0,\rho}(x_{2n}=0)= \mathbb{P}_{0,\varepsilon \cdot|a\rangle\langle a|}(x_{2n}=0) + \mathbb{P}_{0,\rho_\varepsilon }(x_{2n}=0) \geq  \mathbb{P}_{0,\varepsilon \cdot|a\rangle\langle a|}(x_{2n}=0) = 
\varepsilon \cdot \mathbb{P}_{0,|a\rangle\langle a|}(x_{2n}=0).
$$ 
As $\sum_{n\geq 0} \mathbb{P}_{0,|a\rangle\langle a|}(x_{2n}=0) = +\infty$ and $\varepsilon>0$, we conclude $\sum_{n\geq 0} \mathbb{P}_{0,\rho}(x_{2n}=0) = +\infty$.

\medskip

(2) Let $\rho$ be a non-pure density. Hence, there exists an orthonormal basis $\{|v\rangle, |w\rangle\}$  of $\mathbb{C}^2$ and $0<q<1$ such that $\rho = q\cdot|v\rangle\langle v|+(1-q)\cdot|w\rangle\langle w|$.
We have
$$
\mathbb{P}_{0,\rho}(x_{2n}=0)
=q\cdot \mathbb{P}_{0,|v\rangle\langle v|}(x_{2n}=0)
+(1-q)\cdot \mathbb{P}_{0,|w\rangle\langle w|}(x_{2n}=0)
$$
and this implies that
$$
\sum_{n\geq 0} \mathbb{P}_{0,\rho}(x_{2n}=0)
\leq
\sum_{n\geq 0} \mathbb{P}_{0,|v\rangle\langle v|}(x_{2n}=0)
+
\sum_{n\geq 0} \mathbb{P}_{0,|w\rangle\langle w|}(x_{2n}=0).
$$
As $\sum_{n\geq 0} \mathbb{P}_{0,\rho}(x_{2n}=0) = +\infty$, it is not possible
that both 
$$\sum_{n\geq 0} \mathbb{P}_{0,|v\rangle\langle v|}(x_{2n}=0) \;\text{ and }\; \sum_{n\geq 0} \mathbb{P}_{0,|w\rangle\langle w|}(x_{2n}=0)$$
converge. 

\cqfd

\begin{theorem}[Trichotomy for coins of dimension 2]\label{teotri}
Let $(L,R)$ be a coin of dimension 2. We are in one (and only
one) of the following situations:
\begin{itemize}
\item $(L,R)$ is recurrent,
\item $(L,R)$ is transient,
\item $(L,R)$ is recurrent with respect to all densities except for one, which we call $\sigma$. In this case, there exists an eigenvector $|v\rangle$ of both $L$ and $R$ such that $\sigma=|v\rangle\langle v|$.
\end{itemize}
\end{theorem}

\textbf{Proof.} Let $(L,R)$ be a coin of dimension 2 that is neither recurrent nor transient.
 As $(L,R)$ is a coin that is not transient, we have recurrence with respect to some density $\tau$. Lemma \ref{lemrecfaithful}, item 2, allows us to assume that $\tau$ is pure. 
Similarly, as $(L,R)$ is a coin that is not recurrent, we have transience with respect to some density $\sigma$. We will prove that $(L,R)$ is recurrent with respect to all densities but $\sigma$.  

\medskip

Since $(L,R)$ is recurrent with respect to the pure density $\tau$ we have, by Lemma \ref{lemrecfaithful}, item 1, that  $\sigma$ is non-faithful. As the dimension of the coin space is 2, $\sigma$ is pure, so there is a unit vector $|v\rangle$ such that $\sigma=|v\rangle\langle v|.$ Let $|w\rangle$ be a unit vector which is not a multiple of $|v\rangle$. Let $\rho =  q\cdot|v\rangle\langle v|+(1-q)\cdot|w\rangle\langle w|$  with $0<q<1$. Then $\rho$ is non-pure and, as the dimension of the coin space is 2, $\rho$ is faithful. Then, $(L,R)$ is recurrent with respect to $\rho$, by Lemma \ref{lemrecfaithful}, item 1. By combining the recurrence of $\rho$ with the transience of $|v\rangle\langle v|$,  we conclude $(L,R)$ is recurrent with respect to $|w\rangle\langle w|$. Therefore, $(L,R)$ is recurrent with respect to all densities but $\sigma$, by the definition of $|w\rangle$ and by  Lemma \ref{lemrecfaithful}, item 1.  Finally,
$|v\rangle$ must be an eigenvector of $L$ and $R$, as the walk is homogeneous in space and since $(L,R)$ is transient only with respect to $|v\rangle\langle v|$.
\cqfd

The trichotomy of Theorem \ref{teotri} leads to a dichotomy for coins  of  dimension 2 with zero or one common eigenvector. In order to obtain such dichotomy, we slightly modify the result in [\cite{carbone2}, Prop. 6.1], making use of no additional hypotheses over the coin.
\begin{pro}\label{carbreloaded}(Adaptation of [\cite{carbone2}, Prop. 6.1]). Consider a coin $(L,R)$ of dimension 2. Let $\mathfrak{L}$ be the auxiliary map of the induced OQW. Then, we are in one of the following situations:
\begin{enumerate}
\item If $L$ and $R$ have no eigenvector in common, then $\mathfrak{L}$ is irreducible and, as a consequence, has a unique invariant density.
\item If $L$ and $R$ have only one eigenvector $u_1$ in common, then the density $|u_1\rangle\langle u_1|$ is the unique $\mathfrak{L}$-invariant density.
\item  If $L$ and $R$ have two linearly independent eigenvectors in common, then for any nonzero vector $u_2$ orthogonal to one of these two eigenvectors, we have that $u_2$ is also a common eigenvector of $L$ and $R$.
\end{enumerate}
\end{pro}
\textbf{Proof.}
Assume that $\mathfrak{L}$ is not irreducible and that $L$ and $R$ have a common eigenvector $u_1$ having norm 1. Choose $u_2$ so that we have an orthonormal basis $\{u_1,u_2\}$ and let $\rho= \sum_{i,j} \rho_{ij}|u_i\rangle\langle u_j|$ be a $\mathfrak{L}$-invariant density.
We have:
$$
\rho = \mathfrak{L}(\rho) =  \sum_{i,j} \rho_{ij}L|u_i\rangle\langle u_j|L^*+\sum_{i,j} \rho_{ij}R|u_i\rangle\langle u_j|R^*$$
from which we obtain
$$
\rho_{22}= \langle u_2|\mathfrak{L}(\rho)| u_2\rangle = \rho_{22}|\langle u_2|L|u_2 \rangle|^2+\rho_{22}|\langle u_2|R|u_2 \rangle|^2.
$$
Then, we have
\beq\label{adapeq1}
\rho_{22} = 0 
\quad \text{ or } \quad
1= |\langle u_2|L|u_2 \rangle|^2+|\langle u_2|R|u_2 \rangle|^2.
\eeq
Suppose  $\rho_{22} = 0$. As $\rho$ is a density, we have $\rho\geq 0$, which implies $\rho_{12} = 0$ and $\rho = |u_1\rangle \langle u_1|$. Therefore, 
\beq\label{adapeq2}
\rho_{22} = 0 \Rightarrow |u_1\rangle \langle u_1| \text{ is the only invariant density.}
\eeq
Now suppose $\rho_{22} \neq 0$. This implies, $1= |\langle u_2|L|u_2 \rangle|^2+|\langle u_2|R|u_2 \rangle|^2$ and we have 
$$
1-\|L|u_2\rangle\|^2 = \langle u_2 |I| u_2 \rangle - \langle u_2 |L^*L| u_2 \rangle = \langle u_2 |I-L^*L| u_2 \rangle
=\langle u_2 |R^*R| u_2 \rangle = \|R|u_2\rangle\|^2.
$$
By the Cauchy-Schwarz inequality,
$$
\begin{aligned}
|\langle u_2|R|u_2\rangle|^2 &= 1-|\langle u_2|L|u_2\rangle|^2 \\
&\geq 1-\|\langle u_2|\|^2\cdot \|L|u_2\rangle\|^2 \\
&=1-\|L|u_2\rangle\|^2 \\
&= \|R|u_2\rangle\|^2.
\end{aligned}
$$
By Cauchy-Schwarz once again, there exists $\eta \in \mathbb{C}$ such that $R|u_2\rangle = \eta|u_2\rangle$ i.e. $|u_2\rangle$ is an eigenvector of $R$. We can also obtain the same fact for matrix $L$. Therefore,
\beq\label{adapeq3}
1= |\langle u_2|L|u_2 \rangle|^2+|\langle u_2|R|u_2 \rangle|^2 
\Rightarrow 
|u_2\rangle \text{ is a common eigenvector of } L, R.
\eeq
By (\ref{adapeq1}), (\ref{adapeq2}), (\ref{adapeq3}), we have:
\beq\label{adapeq4}
 |u_1\rangle \langle u_1| \text{ is the only invariant density}
\quad \text{ or } \quad
|u_2\rangle \text{ is a common eigenvector of } L, R.
\eeq
Finally, if $L, R$ admit a second common eigenvector $|z\rangle$ linearly independent of $|u_1\rangle$, we have that $|z\rangle\langle z|$ is an invariant density distinct of $ |u_1\rangle \langle u_1|$. By (\ref{adapeq4}),  we have that  $|u_2\rangle$ is a common eigenvector of $L$ and $R$. 
\cqfd

\begin{lemma}\label{lemrecineigen}
Let $(L,R)$ be a coin of finite dimension and let $|v\rangle$ be a common eigenvector of $L$ and $R$. Then
$$
\mathrm{Tr}(L^*L |v\rangle\langle v|) = \frac{1}{2}
 \Leftrightarrow \; 
 (L,R) \text{ is recurrent with respect to }  |v\rangle\langle v|.
$$
\end{lemma}
\textbf{Proof.}
Let $\lambda,\delta \in \mathbb{C}$ such that $L|v\rangle = \lambda|v\rangle$ and $R|v\rangle = \delta|v\rangle$. 
 Let $(x_n,\rho_n)_{n\geq 0}$ be the Markov chain that generates the quantum trajectories of the homogeneous OQW induced by coin $(L,R)$. If  $\rho_0 =  |v\rangle\langle v|$ then $\rho_n=  |v\rangle\langle v|$ for all $n\in\mathbb{N}$, since
\beq\label{neweqredut2}
L|v\rangle\langle v|L^*=|\lambda|^2\cdot |v\rangle\langle v|, \qquad 
R|v\rangle\langle v|R^*=|\delta|^2\cdot |v\rangle\langle v|.
\eeq
Then,
$$
\mathrm{Tr}(L^*L |v\rangle\langle v|) = \langle v|L^*L|v\rangle = |\lambda|^2, \qquad \langle v|R^*R|v\rangle = |\delta|^2.
$$
As $(L,R)$ is a coin, we have that $|\lambda|^2+|\delta|^2 = 1$ and, by (\ref{neweqredut2}), we have 
$
\mathbb{P}_{i,|v\rangle\langle v|}(x_{2n}=i) = \begin{pmatrix} 2n \\ n\end{pmatrix} p^n(1-p)^n
$
with $p=\mathrm{Tr}(L^*L |v\rangle\langle v|) $ and then
 $(L,R)$ is recurrent with respect to $|v \rangle \langle v|$ if, and only if, 
$
\mathrm{Tr}(L^*L |v\rangle\langle v|)  = \frac{1}{2}.
$

\cqfd

\begin{cor}[Dichotomy for coins of dimension 2 with at most one common eigenvector]\label{dicho1eigen}
Let $(L,R)$ be a coin of dimension 2 such that $L$ and $R$ have at most one common eigenvector. We are in one (and only one) of the following situations:
\begin{itemize}
\item $(L,R)$ is transient,
\item $(L,R)$ is recurrent.
\end{itemize}
\end{cor}
\textbf{Proof.}
If $L$ and $R$ have no common eigenvector, the dichotomy follows directly from Theorem \ref{teotri}. If $|v\rangle$ is the unique common eigenvector of both $L$ and $R$, 
suppose that $(L,R)$ is neither transient nor recurrent. Therefore, by  Theorem \ref{teotri}, $(L,R)$ is transient with respect to $\sigma = |v\rangle\langle v|$.
 Hence by Lemma \ref{lemrecineigen} we have
 $\mathrm{Tr}(L^*L \sigma)\neq \frac{1}{2}$. 
By Proposition \ref{carbreloaded}  the auxiliary map $\mathfrak{L}$  has a unique invariant density and, by uniqueness, this density is $\sigma$.  By Proposition \ref{critrecirredcointrans}, $(L,R)$ must be transient, which is a contradiction.
\cqfd

Now we recall that, regarding the recurrence result given by Corollary \ref{critrecirredcoinrec}, the irreducibility of $(L,R)$ was used only to guarantee the uniqueness of the invariant density for the auxiliary map $\mathfrak{L}$, and to ensure that the  dichotomy result of  Proposition \ref{dichot} could be applied. We remark that these two results are also available for coins of dimension 2 with at most one common eigenvector: the former case via Proposition \ref{carbreloaded} and the latter by Corollary \ref{dicho1eigen}. Therefore, Corollary \ref{critrecirredcoinrec} is extendable for  coins of dimension 2 with zero or one common eigenvector:

\begin{pro}\label{atmost1eigen}
Let $(L,R)$ be a coin of dimension 2 with at most one common eigenvector. Let  $\rho_{\infty}$ be the unique invariant density of  the auxiliary map $\mathfrak{L}$. Then,
$$
 \mathrm{Tr}(L^*L\rho_{\infty}) = \frac{1}{2}  \Rightarrow  (L,R) \text{ is recurrent.}
$$
\end{pro}

 Now we examine the case of two common eigenvectors:

\begin{pro}\label{2comeigen}
Let $(L,R)$ be a coin of dimension 2 such that $L$ and $R$ have
 two linearly independent eigenvectors in common. Let $u_1$ be one of them and let $u_2$ be a norm one vector such that $u_2\perp u_1$. Let $\sigma_1 = |u_1\rangle\langle u_1|$ and $\sigma_2 = |u_2\rangle\langle u_2|$.
Then,
$$
\begin{aligned}
 \mathrm{Tr}(L^*L\sigma_1)= \frac{1}{2} \; \text{  and  } \; \mathrm{Tr}(L^*L\sigma_2) = \frac{1}{2} \;  \Rightarrow & \;  (L,R) \text{ is recurrent,}\\
 \mathrm{Tr}(L^*L\sigma_1) \neq \frac{1}{2} \; \text{  and  } \; \mathrm{Tr}(L^*L\sigma_2) \neq \frac{1}{2} \;  \Rightarrow & \; (L,R) \text{ is transient,}\\
\mathrm{Tr}(L^*L\sigma_i) \neq \frac{1}{2} \; \text{  and  } \; \mathrm{Tr}(L^*L\sigma_j) = \frac{1}{2} \;  \Rightarrow &  \;  (L,R) \text{ is transient with respect to } \sigma_i \text{ and it is }\\
&\text{ recurrent with respect to all densities but } \sigma_i, \\ & \text{ for } (i,j) = (1,2) \text{ or } (i,j) = (2,1).\\
\end{aligned}
$$
 
\end{pro}
\textbf{Proof.}
By item 3 of Proposition \ref{carbreloaded}, $u_2$ is also a common eigenvector of $L$ and $R$. Hence $L$ and $R$ are diagonal matrices with respect to the basis $\{u_1,u_2\}$ and $(L,R)$ is unital. Then, if $\mathrm{Tr}(L^*L\sigma_1)= \frac{1}{2} \; \text{  and  } \; \mathrm{Tr}(L^*L\sigma_2) = \frac{1}{2} \;$  then $ (L,R)$  is recurrent  by [\cite{LS2015}, Thm. 4.6]. On the other hand, if   $\mathrm{Tr}(L^*L\sigma_1) \neq \frac{1}{2} \; \text{  and  } \; \mathrm{Tr}(L^*L\sigma_2) \neq \frac{1}{2}$, then by Lemma \ref{lemrecineigen} we have that $(L,R)$ is transient with respect to the two distinct densities $\sigma_1$ and $\sigma_2$, hence $(L,R)$ is transient  by Theorem \ref{teotri}.

\medskip

Finally, if  $\mathrm{Tr}(L^*L\sigma_i) \neq \frac{1}{2} \; \text{  and  } \; \mathrm{Tr}(L^*L\sigma_j) = \frac{1}{2}$  for $(i,j) = (1,2)$ or  $(i,j) = (2,1)$, by Lemma \ref{lemrecineigen} we have that $(L,R)$ is transient with respect to $\sigma_i$ and $(L,R)$ is recurrent with respect to $\sigma_j$. Hence, by Theorem \ref{teotri}, 
$(L,R)$ is transient with respect to $\sigma_i$ and it is recurrent with respect to all densities but  $\sigma_i$.
\cqfd

Summarizing, we gather several of the results about recurrence proved in this section and the previous one in order to obtain a complete set of recurrence criteria for coins of dimension 2.

\begin{theorem}[Recurrence criteria for coins of dimension 2]\label{teocrit}
Consider a homogeneous OQW on the line induced by a coin $(L,R)$ of dimension 2.
\begin{enumerate}
\item If $L$ and $R$ have at most one common eigenvector, let  $\rho_{\infty}$ be the unique invariant density of  the auxiliary map. Then, we have
$$
 \mathrm{Tr}(L^*L\rho_{\infty}) \neq \frac{1}{2}  \Rightarrow  (L,R) \text{ is transient,}
$$
$$
 \mathrm{Tr}(L^*L\rho_{\infty}) = \frac{1}{2}  \Rightarrow  (L,R) \text{ is recurrent.}
$$
\item If $L$ and $R$ have two linearly independent eigenvectors in common, let $u_1$ be one of them and let $u_2$ be a norm one vector such that $u_2\perp u_1$. Also let $\sigma_1 = |u_1\rangle\langle u_1|$ and $\sigma_2 = |u_2\rangle\langle u_2|$. Then, we have
$$
\begin{aligned}
 \mathrm{Tr}(L^*L\sigma_1)= \frac{1}{2} \; \text{  and  } \; \mathrm{Tr}(L^*L\sigma_2) = \frac{1}{2} \;  \Rightarrow & \;  (L,R) \text{ is recurrent,}\\
 \mathrm{Tr}(L^*L\sigma_1) \neq \frac{1}{2} \; \text{  and  } \; \mathrm{Tr}(L^*L\sigma_2) \neq \frac{1}{2} \;  \Rightarrow & \; (L,R) \text{ is transient,}\\
\mathrm{Tr}(L^*L\sigma_i) \neq \frac{1}{2} \; \text{  and  } \; \mathrm{Tr}(L^*L\sigma_j) = \frac{1}{2} \;  \Rightarrow  &  \;  (L,R) \text{ is transient with respect to } \sigma_i \text{ and it is}\\
&\text{  recurrent with respect to all densities but } \sigma_i, \\
& \text{ for } (i,j) = (1,2) \text{ or } (i,j) = (2,1).\\
\end{aligned}
$$

\end{enumerate}
\end{theorem}

\textbf{Proof.} Item 1 is given by  Proposition \ref{critrecirredcointrans} and  Proposition \ref{atmost1eigen}, 
and item 2 is given by Proposition \ref{2comeigen}.
\cqfd

\section{Absorption problem for OQWs on the semi-infinite line}\label{semiinfinite}

Given an OQW on the line, we can always consider an associated OQW on the semi-infinite line if, on eq. (\ref{oqwh1}), we restrict the walk to nonnegative sites $|0\rangle,|1\rangle,\ldots$ and assume that the walk stops after reaching site $|0\rangle$ for the first time (which we may interpret as site $|0\rangle$ being absorbing). Below, recall the definitions of $P_0$ and $Q_0$ given in Section 3.

\medskip


The \textbf{absorption probability on the semi-infinite line} is the probability that an OQW initialized in $\rho \otimes|m\rangle\langle m|$, $m\geq 1$, is eventually absorbed at $0$. This number is
$$
\mathbb{P}_{0\leftarrow m} (\rho) =\sum_{n=0}^\infty Tr[{P}_{0}\Phi({Q}_{0}\Phi)^{n}(\rho \otimes|m\rangle\langle m|)], 
$$
which is essentially the series whose terms are given by (\ref{fprob1}), but regarding an OQW on the semi-infinite line. We say that the OQW is {\bf absorbing at $|0\rangle$}, or simply {\bf absorbing}, if $\mathbb{P}_{0\leftarrow m} (\rho)=1$ for every $\rho$ density and $m\geq 1$.

\subsection{Coins of arbitrary (finite) order}

As an application of the results obtained in previous sections, we will give an absorption criterion for irreducible walks with an internal degree of freedom of arbitrary finite dimension. 

\begin{pro}\label{absorothercoinsprouniqueinvst} 
Let $(L,R)$ be a coin inducing an OQW and an absorbing walk on the semi-infinite line, where $L$ is the transition matrix to the left. Assume that the auxiliary map admits a unique invariant state  $\rho_{\infty}$. Then,  
$$
\mathrm{Tr}(L^*L\rho_{\infty}) > \frac{1}{2}
 \; \Rightarrow   \;
\text{The OQW is absorbing}.
$$
\end{pro}
\textbf{Proof.}
 Let $(x_n,\rho_n)_{n\geq 0}$ be the Markov chain that generates the quantum trajectories of the OQW spanned by coin $(L,R)$.
By [\cite{attalhom}, Thm. 5.2], the law of large numbers $\lim_{n \to \infty} \frac{x_n}{n} = \mu$ holds almost surely, where $\mu = \mathrm{Tr}(R\rho_{\infty}R^*)-\mathrm{Tr}(L\rho_{\infty}L^*) = 1 -2 \mathrm{Tr}(L^*L\rho_{\infty})<0$.
As $\mu <0$, we have $x_n  \rightarrow -\infty\; a.s.$ which clearly implies that the probability of return to $0$ equals 1.
Therefore,
$$
\begin{array}{ll}
\mathbb{P}_{0\leftarrow m} ( \rho) = 1  & \text{ if } \mathrm{Tr}(L^*L\rho_{\infty}) > \frac{1}{2}, \\
\end{array}
$$
for any density matrix  $\rho$   and $m\geq 1$. 
\cqfd

\begin{pro}\label{absorothercoinsproirred} 
Let $(L,R)$ be a coin inducing an irreducible OQW and an absorbing walk on the semi-infinite line, where $L$ is the transition matrix to the left.  Let  $\rho_{\infty}$ be the unique invariant state of  the auxiliary map. Then,  
$$
\begin{aligned}
\mathrm{Tr}(L^*L\rho_{\infty}) = \frac{1}{2}
 &\; \Rightarrow   \;
\text{The OQW is absorbing},
\\
\mathrm{Tr}(L^*L\rho_{\infty}) < \frac{1}{2}
&\; \Rightarrow   \;
\text{The OQW is not absorbing}.
\end{aligned}
$$
\end{pro}
\textbf{Proof.} Suppose $\mathrm{Tr}(L^*L\rho_{\infty}) = \frac{1}{2}.$ By Corollary \ref{critrecirredcoinrec}, the walk is recurrent. As it is also irreducible, the probability to be eventually absorbed in 0 equals 1, by Proposition \ref{equivhom}.

\medskip

Suppose $\mathrm{Tr}(L^*L\rho_{\infty}) < \frac{1}{2}.$ Let $(x_n,\rho_n)_{n\geq 0}$ be the Markov chain that generates the quantum trajectories of the OQW spanned by coin $(L,R)$.
By [\cite{attalhom}, Thm. 5.2], the law of large numbers $\lim_{n \to \infty} \frac{x_n}{n} = \mu$ holds almost surely, where $\mu = \mathrm{Tr}(R\rho_{\infty}R^*)-\mathrm{Tr}(L\rho_{\infty}L^*) = 1 -2 \mathrm{Tr}(L^*L\rho_{\infty})>0$.
As $\mu>0$, we have $x_n  \rightarrow +\infty\; a.s.$ which  implies that the mean number of visits to $0$ is finite, so that the monitored probability of return to $0$ is less than 1, by Proposition \ref{equivhom}.
Therefore,
$$
\left\{
\begin{array}{ll}
\mathbb{P}_{0\leftarrow m} ( \rho) = 1  & \text{ if } \mathrm{Tr}(L^*L\rho_{\infty}) = \frac{1}{2}, \\
\mathbb{P}_{0\leftarrow m} ( \rho) < 1 & \text{ if } \mathrm{Tr}(L^*L\rho_{\infty}) < \frac{1}{2}, \\
\end{array}
\right.
$$
for any density matrix  $\rho$   and $m\geq 1$. 
\cqfd

\begin{theorem}[Absorption criterion for irreducible walks on the semi-infinite line]\label{absorothercoins} 
Let $(L,R)$ be a coin inducing an irreducible OQW and an absorbing walk on the semi-infinite line, where $L$ is the transition matrix to the left.  Let  $\rho_{\infty}$ be the unique invariant state of  the auxiliary map. Then,  
$$
\mathrm{Tr}(L^*L\rho_{\infty}) \geq \frac{1}{2}
 \; \Leftrightarrow   \;
\text{The OQW is absorbing}.
$$
\end{theorem}
\textbf{Proof.}
The case of $\mathrm{Tr}(L^*L\rho_{\infty}) > \frac{1}{2} $ is given by Proposition \ref{absorothercoinsprouniqueinvst} 
and the case of $\mathrm{Tr}(L^*L\rho_{\infty}) \leq \frac{1}{2} $ is given by Proposition \ref{absorothercoinsproirred}. 
\cqfd
 

\subsection{Coins of order 2}

In order to prove the absorption criterion  for coins of dimension 2, we need two auxiliary lemmas.

\begin{lemma}\label{lemabsoreigen}
Consider the homogeneous OQW  on the semi-infinite line induced by 
a coin $(L,R)$ of dimension $2$, where $L$ is the transition matrix to the left. Suppose there exists a common eigenvector for  $R$ and $L$, which we call $|v\rangle$. Then, 
$$
\mathrm{Tr}(L^*L |v\rangle\langle v|) \geq \frac{1}{2}
 \Leftrightarrow \; 
 \mathbb{P}_{0\leftarrow m} ( |v\rangle\langle v|) = 1,\;\;\;m\geq 1.
$$
\end{lemma}
\textbf{Proof.}
Let $\lambda,\delta \in \mathbb{C}$ such that $R|v\rangle = \lambda|v\rangle$ and $L|v\rangle = \delta|v\rangle$. 
 Let $(x_n,\rho_n)_{n\geq 0}$ be the Markov chain that generates the quantum trajectories of the OQW spanned by coin $(L,R)$. If  $\rho_0 =  |v\rangle\langle v|$ then $\rho_n=  |v\rangle\langle v|$ for all $n\in\mathbb{N}$, since
\beq\label{neweqredut2abs}
R|v\rangle\langle v|R^*=|\lambda|^2\cdot |v\rangle\langle v|, \qquad 
L |v\rangle\langle v|L^*=|\delta|^2\cdot |v\rangle\langle v|
\eeq
and this implies
$$
\mathrm{Tr}(R^*R |v\rangle\langle v|) = \langle v|R^*R|v\rangle = |\lambda|^2, \qquad \langle v|L^*L|v\rangle = |\delta|^2.
$$
As $(L,R)$ is a coin, we have $|\lambda|^2+|\delta|^2 = 1$ and, by (\ref{neweqredut2abs}), we have that this OQW behaves as a classical, simple random walk, with $|\lambda|^2$ being the probability of moving right and $|\delta|^2$ the probability of moving left. Hence, we have
$$
 |\delta|^2 \geq \frac{1}{2}
 \Leftrightarrow \; 
 \mathbb{P}_{0\leftarrow m} ( |v\rangle\langle v|) = 1.
$$
\cqfd

\begin{lemma}\label{lemaTr0}
Consider the OQW  on the semi-infinite line induced by a coin $(L,R)$ such that $R$ and $L$ are diagonal matrices of dimension 2, and $L$ is the transition matrix to the left. Let 
$\tau = \begin{bmatrix} 0 & \zeta \\ \xi & 0\end{bmatrix}$ with $\zeta,\xi \in \mathbb{C}$. Then $ \mathbb{P}_{0\leftarrow m} (\tau) = 0$.
\end{lemma}
\textbf{Proof.} This follows from the fact that $\mathrm{Tr}(R^*R\tau) = \mathrm{Tr}(L^*L\tau) = 0$ and noting that, as $L$ and $R$ are diagonal, so are $L^*L$ and $R^*R$. 
\cqfd

\begin{theorem}[Absorption criterion for coins of dimension 2 on the semi-infinite line]\label{absorcoinsdim2} Let $(L,R)$ be a coin of dimension $2$ inducing an OQW and an absorbing walk on the semi-infinite line, where $L$ is the transition matrix to the left. 
\begin{enumerate}
\item If $R$ and $L$ have at most one common eigenvector, let  $\rho_{\infty}$ be the unique invariant density of  the auxiliary map. Then,
$$ \text{The OQW is absorbing} \;\Leftrightarrow\; \mathrm{Tr}(L^*L\rho_{\infty}) \geq \frac{1}{2}.$$

\item If $R$ and $L$ have two linearly independent eigenvectors in common, let $u_1$ be one of them and let $u_2$ be a norm one vector such that $u_2\perp u_1$. Let $\sigma_1 = |u_1\rangle\langle u_1|$ and $\sigma_2 = |u_2\rangle\langle u_2|$. Then,
$$
\begin{aligned}
\text{The OQW is absorbing} \; \Leftrightarrow \;
 \mathrm{Tr}(L^*L\sigma_1)\geq \frac{1}{2} \; \text{  and  } \; \mathrm{Tr}(L^*L\sigma_2) \geq \frac{1}{2}.\end{aligned}
$$
\end{enumerate}
\end{theorem}

\textbf{Proof.} (1) If $\mathrm{Tr}(L^*L\rho_{\infty}) = \frac{1}{2}$, we have that the walk is recurrent by Theorem \ref{teocrit}, item 1. Now, let us prove that $\mathbb{P}_{0,\rho}(t_0<\infty)=1$ for every $\rho$, i.e., that the walk is absorbing.
Since $\mathbb{E}_{0,\rho}(n_0)=\infty$ for every $\rho$, by item (4) in Section 3.1, we have that $\mathbb{P}_{0,\rho'}(t_0<\infty)=1$ for some $\rho'$ which is accessible from $\rho$. If $\rho$ is the pure state given by the common eigenvector of $L$ and $R$ then $\rho'=\rho$, so we have a classical walk and absorption holds. On the other hand, if $\rho$ is not such common eigenstate, we can find $\rho'\neq\rho$ for which the walk is also absorbing. Therefore, by any nontrivial convex combination of such densities one can obtain an absorbing density which is faithful and conclude, by item (2) in Section 3.1, that the walk is absorbing.

\medskip

Now suppose $\mathrm{Tr}(L^*L\rho_{\infty}) \neq \frac{1}{2}$. The law of large numbers $\lim_{n \to \infty} \frac{x_n}{n} = \mu$ holds almost surely, where $\mu = \mathrm{Tr}(R\rho_{\infty}R^*)-\mathrm{Tr}(L\rho_{\infty}L^*) = 1 -2 \mathrm{Tr}(L^*L\rho_{\infty})$ and as $\mu \neq 0$, we have
$$
\left\{
\begin{array}{ll}
x_n  \rightarrow +\infty\; a.s. & \text{ if } \mu > 0, \\
x_n  \rightarrow -\infty\; a.s. & \text{ if } \mu < 0 .\\
\end{array}
\right.
$$
The case $\mu<0$ clearly implies that the walk is absorbing. On the other hand, the case $\mu> 0$ implies that the mean number of visits to $0$ is finite, for every initial $\rho$. By [\cite{bbp}, Cor. 3.5], for every faithful state $\rho$, we must have $\mathbb{P}_{0,\rho}(t_0<\infty)<1$. Moreover, the same must hold for every pure state. In fact, 
suppose there is a pure state $\eta$ such that $\mathbb{P}_{0,\eta}(t_0<\infty)=1$.
If $\eta$ is the pure state given by the common eigenvector of $L$ and $R$ then  we have a classical walk and 
we conclude that the mean number of visits to $0$ is infinite, a contradiction. On the other hand, if $\eta$ is not such common eigenstate, then at some time the walk must land at some density $\eta'\neq\eta$ at site $0$. As $\eta'$ is accessible from $\eta$, for which the walk is absorbing, we must have $\mathbb{P}_{0,\eta'}(t_0<\infty)=1$, by a similar reasoning as for Proposition \ref{equivhom}. Then, any nontrivial convex combination of $\eta$ and $\eta'$ would be a faithful density $\gamma$ with $\mathbb{P}_{0,\gamma}(t_0<\infty)=1$, a contradiction by item (2) in Section 3.1. Hence, the case $\mu> 0$ implies that the walk is not absorbing. Therefore,
$$
\left\{
\begin{array}{ll}
\mathbb{P}_{0\leftarrow m} ( \rho) < 1 & \text{ if } \mathrm{Tr}(L^*L\rho_{\infty}) < \frac{1}{2}, \\
\mathbb{P}_{0\leftarrow m} ( \rho) = 1  & \text{ if } \mathrm{Tr}(L^*L\rho_{\infty}) \geq \frac{1}{2}, \\
\end{array}
\right.
$$
for any density matrix  $\rho$ and $m\geq 1$.

\medskip

(2) By Proposition \ref{carbreloaded}, $|u_2\rangle$ is also an eigenvector of $L$ and $R$, which implies that $(L,R)$ is a diagonal coin with respect to basis $\{|u_1\rangle,|u_2\rangle\}$. By Lemma \ref{lemabsoreigen}, we have:
$$
  \mathrm{Tr}(L^*L\sigma_1) \geq \frac{1}{2} \Leftrightarrow   \mathbb{P}_{0\leftarrow m} ( \sigma_1) = 1,\;\;\;m\geq 1,
$$
$$
  \mathrm{Tr}(L^*L\sigma_2) \geq \frac{1}{2} \Leftrightarrow   \mathbb{P}_{0\leftarrow m} ( \sigma_2) = 1,\;\;\;m\geq 1.
$$
For $\rho = \sum_{1\leq i,j \leq 2}\rho_{ij}|u_i\rangle \langle u_j|$, we can write
$$
\mathbb{P}_{0\leftarrow m} (\rho)=
\rho_{11}\mathbb{P}_{0\leftarrow m} ( \sigma_1)
+\rho_{12}\mathbb{P}_{0\leftarrow m} (|u_1\rangle \langle u_2|)
+\rho_{21}\mathbb{P}_{0\leftarrow m} (|u_2\rangle \langle u_1|)
+\rho_{22}\mathbb{P}_{0\leftarrow m} ( \sigma_2)
$$
$$
=\rho_{11}\mathbb{P}_{0\leftarrow m} ( \sigma_1)
+\rho_{22}\mathbb{P}_{0\leftarrow m} ( \sigma_2),
$$
the latter equality by Lemma \ref{lemaTr0}. Since $\rho_{11}+\rho_{22} = 1$ we have, for every $m\geq 1$,
$$
\left(
 \mathrm{Tr}(L^*L\sigma_1)\geq \frac{1}{2} \; \text{  and  } \; \mathrm{Tr}(L^*L\sigma_2) \geq \frac{1}{2}
\right)
 \;  \Leftrightarrow  \;   
\left(
  \mathbb{P}_{0\leftarrow m} ( \sigma_1) = 1  \; \text{  and  } \;    \mathbb{P}_{0\leftarrow m} ( \sigma_2) = 1
\right)
 \;  \Leftrightarrow  \;  $$
 $$
 \;  \Leftrightarrow  \; \mathbb{P}_{0\leftarrow m} ( \rho) = 1.
$$

\cqfd

\section{Examples}\label{sec7}

\begin{example}[PQ-matrices]\label{transientwithTrBB1} A PQ-channel is a completely positive map which admits a Kraus decomposition $\sum_i V_i\cdot V_i^*$ such that each $V_i$ is a permutation of a diagonal matrix. We call such matrices PQ-matrices, see \cite{LS2015}. 

\medskip

a) Consider the OQW induced by the diagonal coin $(L,R)$ with 
$$L=\begin{bmatrix} \frac{1}{\sqrt{3}} & 0 \\ 0 & \frac{\sqrt{2}}{\sqrt{3}} \end{bmatrix},\;\;\; R= \begin{bmatrix} \frac{\sqrt{2}}{\sqrt{3}} & 0 \\ 0 & \frac{1}{\sqrt{3}} \end{bmatrix}.$$
It holds that $L$ and $R$ have $e_1=[1\;0]^T$ and $e_2=[0\; 1]^T$ as eigenvectors, and a calculation leads to $\mathrm{Tr}(L^* L |e_1\rangle\langle e_1|)=1/3$ and $\mathrm{Tr}(L^* L |e_2\rangle\langle e_2|)=2/3$. Therefore, the OQW induced by such coin is transient by part 2 of Theorem \ref{teocrit}. We note that this result can also be derived by using [\cite{LS2015}, Thm. 4.6]. 

\medskip

b) The authors of \cite{LS2015} have proved a recurrence criterion for homogeneous OQWs on the line with a coin $(L,R)$ consisting of PQ-matrices, and such that the auxiliary map is unital. With the results of the present work, site-recurrence of non-unital examples can be easily examined as well: consider the following variation of item a), given by
$$L=\begin{bmatrix} \frac{1}{\sqrt{3}} & 0 \\ 0 & \frac{\sqrt{2}}{\sqrt{3}} \end{bmatrix},\;\;\; R= \begin{bmatrix} 0 & \frac{1}{\sqrt{3}} \\ \frac{\sqrt{2}}{\sqrt{3}} & 0 \end{bmatrix}.$$
We have that
$$\rho_{\infty}= \frac{1}{3}\cdot 
\begin{bmatrix} 1 & 0 \\ 0 & 2 \end{bmatrix}$$
is the unique invariant state of $\mathfrak{L}$ and $ \mathrm{Tr}(L^*L\rho_{\infty})=\frac{5}{9}>  \frac{1}{2}$. Therefore, the irreducible OQW induced by such coin is transient by part 1 of Theorem \ref{teocrit}. As for absorption, by Theorem \ref{absorothercoins} we have, with respect to the OQW on the semi-infinite line induced by such coin, that $\mathbb{P}_{0\leftarrow m} ( \rho) = 1$, for any density and $m\geq 1$.

\medskip

c) Consider the coin $(L,R)$ given by:
$$
L= \begin{bmatrix} \frac{1}{\sqrt{3}} & 0\\ 0 &  \frac{1}{\sqrt{2}} \end{bmatrix} ,
\qquad
R= \begin{bmatrix} \frac{\sqrt{2}}{\sqrt{3}}   & 0 \\ 0 &  \frac{1}{\sqrt{2}}  \end{bmatrix} .
$$
Let us discuss both recurrence and absorption of a site.

\medskip

{Recurrence:} the OQW induced by such coin is transient with respect to density $|e_1\rangle\langle e_1|$ and it is recurrent
for all densities $\rho \neq |e_1\rangle\langle e_1|$.
In fact, the canonical vectors $|e_1\rangle, |e_2\rangle$ 
are both eigenvectors of $R$ and $L$. If we set $\sigma_1 = |e_1\rangle\langle e_1|$ and $\sigma_2 = |e_2\rangle\langle e_2|$, we have
$ \mathrm{Tr}(L^*L\sigma_1)= \frac{1}{3} \; \text{  and  } \; \mathrm{Tr}(L^*L\sigma_2) = \frac{1}{2}$ and, 
by Theorem \ref{teocrit}, we obtain the result.

\medskip

{Absorption:} we have 
$
\mathbb{P}_{0\leftarrow m}(|e_2\rangle\langle e_2|) = 1
$ 
 and
$
\mathbb{P}_{0\leftarrow m}(\rho) < 1
$ 
for all densities $\rho \neq |e_2\rangle\langle e_2|$.
In fact, the canonical vectors $|e_1\rangle, |e_2\rangle$ 
are both eigenvectors of $R$ and $L$. If we set $\sigma_1 = |e_1\rangle\langle e_1|$ and $\sigma_2 = |e_2\rangle\langle e_2|$, we have
$ \mathrm{Tr}(L^*L\sigma_1)= \frac{1}{3} \; \text{  and  } \; \mathrm{Tr}(L^*L\sigma_2) = \frac{1}{2}$ and,
by Theorem \ref{absorcoinsdim2}, we are done.
\end{example}
\qee

\begin{example}[Unitary sum coins]\label{excoin03} Let   $a,b \in \mathbb{C}\backslash \{0\}$ and
$$
L = \begin{bmatrix} a & b \\  0 & 0 \end{bmatrix} ,
\quad
R=\begin{bmatrix} 0 & 0 \\  -\bar{b} & \bar{a}\end{bmatrix} .
$$
Note that  the pair $(L,R)$ is a coin if, and only if, $|a|^2+|b|^2=1$. We have that the OQW induced by such coin is recurrent, by Theorem \ref{teocrit}, since the walk is irreducible with $\rho_\infty=I/2$, thus implying $\mathrm{Tr}(L^*L\rho_\infty)=1/2$. 
\end{example}
\qee

\begin{example}
[Unbalanced coin] Let
$$
L =\begin{bmatrix} \frac{1}{\sqrt{3}} & \frac{1}{\sqrt{2}} \\ \frac{1}{\sqrt{3}} & 0 \end{bmatrix},
\quad
R =\begin{bmatrix} \frac{1}{\sqrt{3}} & -\frac{1}{\sqrt{2}} \\ 0 & 0 \end{bmatrix}.
$$
In \cite{JL}, the authors have proved the positive recurrence of the OQW on the semi-infinite line induced by such coin via an application of Foster's Theorem for OQWs. In particular, this implies that the absorption probability of the origin equals 1. Theorem \ref{absorothercoins} provides a simpler, alternative proof of this fact. Indeed, the induced OQW is irreducible and we have that 
$$\rho_{\infty}= \frac{1}{20}\cdot 
\begin{bmatrix} 15 & 6+\sqrt{6} \\ 6+\sqrt{6} & 5 \end{bmatrix}$$
is the unique invariant state of the auxiliary map. Then, we have:
$$
 \mathrm{Tr}(L^*L\rho_{\infty})=  \frac{29}{40}+\frac{\sqrt{6}}{10}>\frac{1}{2}.
$$
Therefore, for any density $\rho$  and $m\geq 1$ we conclude, by Theorem \ref{absorothercoins}, that $\mathbb{P}_{0\leftarrow m} ( \rho) = 1$. We also conclude the transience of the OQW on the line induced by such coin, by Theorem \ref{teocrit}.
\eex
\cqfde

\bex Let
$$L=\frac{1}{\sqrt{3}}\cdot\begin{bmatrix} 1 & 1 \\ 0 & 1\end{bmatrix},\;\;\;R=\frac{1}{\sqrt{3}}\cdot\begin{bmatrix} 1 & 0 \\ -1 & 1\end{bmatrix}.$$
The homogeneous OQW induced by this coin was described in \cite{attal}, and in [\cite{cgl}, Sec. 8] its site-recurrence was proved by studying the spectrum of the Fourier transform of such walk. With the results presented in this work, we can obtain a straightforward proof of such recurrence: noting that $L$ and $R$ have no eigenvectors in common, and that the auxiliary map is unital, we have
$$\mathrm{Tr}(L^*L\rho_\infty)=\frac{1}{2},\;\;\;\rho_\infty=\frac{I}{2}.$$
Therefore, with respect to the homogeneous OQW on the line, site recurrence holds with respect to every density, by Theorem \ref{teocrit} and, with respect to the walk on the semi-infinite line, we have that the origin is absorbing, by Theorem \ref{absorothercoins}.
\eex
\qee

\begin{example}\label{exampleTrB*Bneq1} Consider the class of non-unital coins given by
$$ 
L=\begin{bmatrix} \frac{1}{\sqrt{2}} & x \\ 0 & \frac{1}{\sqrt{2}}\sqrt{1-2x^2}\end{bmatrix} ,
\qquad
R=\begin{bmatrix} \frac{1}{\sqrt{2}} & -x \\ 0 & \frac{1}{\sqrt{2}}\sqrt{1-2x^2}\end{bmatrix},\;\;\;0<x<\frac{1}{2}.
$$
Then, the OQW induced by $(L,R)$ is recurrent. In fact, we have that $|e_1\rangle$ is the unique common eigenvector of $L$ and $R$, $|e_1\rangle\langle e_1|$ is the unique invariant density for the auxiliary map and
$ \mathrm{Tr}(R^*R|e_1\rangle\langle e_1|) = \frac{1}{2}$. Therefore, by Theorem \ref{teocrit}, we obtain the result.

\end{example}
\cqfde

\begin{example} Let
$$L=\frac{1}{3}\begin{bmatrix} \sqrt{6} & 0 & 0 \\ 0 & 0 & \sqrt{3} \\ -1 & \sqrt{6} & 0\end{bmatrix},\;\;\;R=\frac{1}{3}\begin{bmatrix} 0 & 0 & \sqrt{6} \\ 1 & \frac{\sqrt{6}}{2} & 0 \\ -1 & -\frac{\sqrt{6}}{2} & 0 \end{bmatrix}$$
It is a simple matter to show that the auxiliary map has exactly one invariant density $\rho_\infty$, and a simple calculation leads to
$$\mathrm{Tr}(L^*L\rho_\infty)\approx 0.717825,$$
from which we conclude that the homogeneous OQW induced by $(L,R)$ is transient, by Proposition \ref{critrecirredcointrans}. By Proposition \ref{absorothercoinsprouniqueinvst}, if $L$ is the transition matrix to the left, we conclude the walk on the semi-infinite line is absorbing.

\end{example}
\cqfde

{\bf \noindent Acknowledgements} 

\noindent The authors are grateful to the anonymous referees for their careful reading of the work. We also  would like to thank R. Carbone and Y. Pautrat for a discussion regarding their work on OQWs. TSJ acknowledges financial support by CAPES (Coordena\c c\~ao de Aperfei\c coamento de Pessoal de N\'ivel Superior) during the period 2017-2018 and CNPq (Conselho Nacional de Desenvolvimento Cient\'ifico e Tecnol\'ogico), during the period 2018-2020 (PPGMat/UFRGS). CFL acknowledges financial support from a CAPES/PROAP grant (Programa de Apoio \`a P\'os-Gradua\c c\~ao - 2019) to PPGMat/UFRGS.

\nonumsection{References}

\end{document}